\RequirePackage{fix-cm}
\documentclass[twocolumn]{svjour3}          % twocolumn
\smartqed  % flush right qed marks, e.g. at end of proof
\usepackage{graphicx}
\usepackage[misc,geometry]{ifsym}
\usepackage{natbib}

\bibpunct{(}{)}{;}{a}{}{,}
\usepackage{times}
\usepackage{amssymb}
\usepackage{amsmath}
\usepackage{multirow}
\usepackage{tabularx}
\usepackage{booktabs}
\usepackage{setspace}
\usepackage{rotating}
\usepackage{array,graphicx,subfig}
\usepackage{lscape}
\newcommand{\tabincell}[2]{\begin{tabular}{@{}#1@{}}#2\end{tabular}}

\newcommand{\etal}{\textit{et al}.}
\journalname{Journal of Geodesy}

\begin{document}

\title{An analytical method for error analysis of GRACE-like missions based on spectral
analysis
%\thanks{Grants or other notes
%about the article that should go on the front page should be
%placed here. General acknowledgments should be placed at the end of the article.}
}
%\subtitle{Do you have a subtitle?\\ If so, write it here}

%\titlerunning{Short form of title}        % if too long for running head

\author{
    Lin Cai  \and
    Zebing Zhou \and
    Qiong Li \and
    Zhicai Luo \and
    Houtse Hsu
    }

%\authorrunning{Short form of author list} % if too long for running head
\institute{L. Cai \and Z. Zhou (\Letter) \and Q. Li \and Z. Luo
            \at MOE Key Laboratory of Fundamental Physical Quantities Measurement, School of Physics, Huazhong University of Science and Technology, Wuhan 430074, China\\
              \email{zhouzb@hust.edu.cn}
        \and
            L. Cai \and Z. Zhou \and Q. Li \and Z. Luo \and H. Hsu
            \at Institute of Geophysics, Huazhong University of Science and Technology, Wuhan 430074, China
\\
            \email{cailin@hust.edu.cn}
            \and
            H. Hsu
            \at Institute of Geodesy and Geophysics (IGG), Chinese Academy of Sciences, Wuhan 430074, China
            %Tel.: +86-027-87556651\\
            %Fax: +86-027-678910\\
             %Tel.: +123-45-678910\\
              %Fax: +123-45-678910\\
              %\email{fauthor@example.com}           %  \\
%             \emph{Present address:} of F. Author  %  if needed
}

\date{Received: date / Accepted: date}
% The correct dates will be entered by the editor

\maketitle

\begin{abstract}
The aim of this paper is to present an analytical relationship between the power spectral
density of GRACE-like mission measurements and the accuracies of the gravity field
coefficients mainly from the point of view of theory of signal and system, which indicates
the one-to-one correspondence between spherical harmonic error degree variances and
frequencies of the measurement noise. In order to establish this relationship, the average
power of the errors due to gravitational acceleration difference and the relationship
between perturbing forces and range-rate perturbations are derived, based on the
orthogonality property of associated Legendre functions and the linear orbit perturbation
theory, respectively. This method provides a physical insight into the relation between
mission parameters and scientific requirements. By taking GRACE-FO as the object of
research, the effects of sensor noises and time variable gravity signals are analyzed. If
LRI measurements are applied, a mission goal with a geoid accuracy of 7.4 cm at a spatial
resolution of 101 km is reachable, whereas if the KBR measurement error model is applied, a
mission goal with a geoid accuracy of 10.2 cm at a spatial resolution of 125 km is
reachable. Based on the discussion of the spectral matching of instrument accuracies, an
improvement in accuracy of accelerometers is necessary for the match between the range
errors and accelerometer noises in the future mission. Temporal aliasing caused by the time
variable gravity signals is also discussed by this method.

\end{abstract}

\section{Introduction}
\label{intro} The last dedicated gravity satellite missions like CHAMP, GRACE, GOCE and GRAIL
have mapped the Earth's and Moon's gravity field with unprecedented high accuracy and
resolution in the past decades \citep{Reigber 2002,Tapley 2004,Rummel 2011,Zuber 2013}. CHAMP
and GOCE are mainly based on satellite-to-satellite tracking in the high-low mode (HL-SST)
and satellite gravity gradiometry (SGG) respectively, while both GRACE and GRAIL
satellite-to-satellite tracking use the low-low mode (LL-SST). Compared to HL-SST and SGG
configurations, the LL-SST observations can derive the long wavelength components of the
Earth's gravity field with higher accuracy and map their variability in time in an efficient
way. LL-SST missions based on intersatellite ranging may achieve significant improvements in
spatial resolution and accuracy of gravity field model by using interferometric laser ranging
instead of microwave ranging. Due to these advantages, the proposed future missions, like the
GRACE Follow-On (GRACE-FO) \citep{Flechtner 2015}, Next-Generation Gravity Mission (NGGM)
\citep{Cesare and Sechi 2013} concept and Earth System Mass Transport Mission (e.motion)
proposal \citep{Gruber 2014}, are all based on LL-SST configuration. Until now there exist
four basic types of LL-SST satellites formations for the missions to choose from, i.e.
collinear tandem (GRACE-like), pendulum, Cartwheel and LISA-type formation
\citep[c.f.][]{Elsaka 2014}. Several studies were published to investigate the performance of
these satellite formations, e.g. by \citet{Sharifi 2007}, \citet{Sneeuw 2008}, \citet{Wiese
2009}, \citet{Massotti 2013}, \citet{Elsaka 2014} and \citet{Elsaka 2015}.

The upcoming GRACE-FO mission based on the collinear tandem configuration is about to be
launched in 2017 and will have a nominal life-time of 7 years \citep{Flechtner 2014}. By
taking advantages of GRACE and GRAIL heritage, the GRACE-FO mission will continue to obtain
the global models of the Earth's time-variable gravity field, while on the other hand it will
try to improve the LL-SST measurement performances. For this purpose, a 50-100 nm precise
laser ranging interferometer (LRI) is included into the GRACE-FO payload as a science
demonstrator instrument, which supplements the $\mu $m-level accuracy K-band ranging system
(KBR). The GRACE-FO mission is expected to provide meaningful guidance to the future gravity
satellite missions of LL-SST type after GRACE-FO.

The pre-mission error analysis is a key issue for the future mission design, which concerns
the field where geodesy is in contact with physics and technical sciences. It allows one to
determine the science requirements and parameters of missions before launch. The conventional
error analysis and recovery methods of LL-SST are based on orbit perturbation theory or the
principle of energy conservation in establishing the observation equations, which are
generally solved by using least-squares (LS) theory \citep{Colombo 1984,Touboul 1999,Tapley
2004}. However, there was no one-to-one correspondence between spherical harmonics and
frequencies in the measurements \citep{Inacio 2015}, i.e. accelerometer data, range-rate
data. That means the conventional methods estimate the individual effects of parameters and
noise are too complicated to be described analytically since these methods address the effect
of measurement errors mainly from a numerical point of view \citep{Migliaccio 2004,Cai 2012}.

By applying the theory of signal and system, this paper provides an analytical relationship
between the power spectral density (PSD) of LL-SST measurements and the accuracies of gravity
field coefficients, which indicates the one-to-one correspondence between spherical harmonic
error degree variances and frequencies of the measurement noise. This error analysis method
allows us to efficiently evaluate the science requirements and parameters of the missions. It
is a helpful tool for identifying the frequency characteristics of signals in future gravity
missions.

\citet{Sneeuw 2000} and \citet{Kim 2000} developed their respective semi-analytical theory on
error analysis with different principles. The semi-analytical approach established by
\citet{Sneeuw 2000} obtains the 2-D Fourier spectrum first by Fourier analysis and then
transforms the Fourier coefficients into the spherical harmonic coefficients. In the latter
step, the relationship between spherical harmonics and 2-D Fourier spectrum cannot be
analytically given and must be preceded with applying least-squares. The semi-analytical
method for degree error prediction established by \citet{Kim 2000} can obtain degree error
variance of the gravity by a expression when that of range-rate is available. But before this
step, the range-rate measurement noises due to various error sources need to be covered the
entire sphere with the same latitude and longitude lengths and then mapped from the space
domain into the spectral domain to obtain the degree variance of range-rate. These works made
a significant contribution to the progress of efficient computation of error analysis of
gravity field, however, these methods cannot lead to directly evaluate the frequency
characteristics of measurement noise which affects spherical harmonic coefficient recovery
due to the lack of analytical expression.

This paper, with GRACE-FO as the object of the research, discusses an analytical error
analysis method of LL-SST (a collinear tandem configuration), and is organized as follows. In
Sect. 2, the forces variation relationship between two satellites produced by the
gravitational and non-gravitational accelerations is derived based on dynamic analysis of the
satellite. The information of the range-rate is put in relation with the differential effect
of the resultant forces acting on the twin satellites, which consist of gravitational terms,
due to the gravity field of the Earth and third bodies, and non-gravitational terms, due to
the surface forces like atmospheric drag and solar radiation. A direct analytical expression
for the error analysis of LL-SST is then concluded based on the dynamic analysis and spectral
analysis in Sect. 3. The transfer function between satellite perturbing forces and the
range-rate are deduced in detail in Sects. 4. In Sect. 5, the effects of sensor noise and
their matching together with temporal aliasing from both non-tidal and tidal sources on
gravity field recovery are explicitly and quantitatively discussed by taking the advantage of
this method.

\section{Dynamic analysis}
\label{sec:2} The fundamental relation of the LL-SST is the forces variation between two
satellites produced by the gravitational and non-gravitational accelerations, which can be
expressed in the inertial frame
\begin{eqnarray} \label{eq1}
\mathrm{\Delta} \mathord{\buildrel{\lower3pt\hbox{$\scriptscriptstyle\rightharpoonup$}}
\over a}  = \mathrm{\Delta} \mathord{\buildrel{\lower3pt\hbox{$\scriptscriptstyle\rightharpoonup$}}
\over g}  + \mathrm{\Delta} {\mathord{\buildrel{\lower3pt\hbox{$\scriptscriptstyle\rightharpoonup$}}
\over a} _{ng}},
\end{eqnarray}
\begin{figure}
\begin{center}
\centering
% Use the relevant command to insert your figure file.
% For example, with the graphicx package use
  \includegraphics[angle=0, width=0.45\textwidth]{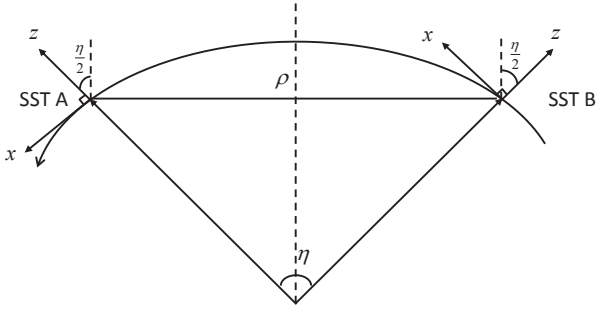}
% figure caption is below the figure
\caption{Configuration of two collinear tandem satellites}
\label{fig:1}       % Give a unique label
\end{center}
\end{figure}
where $\mathrm{\Delta}\mathord{\buildrel{\lower3pt\hbox{$\scriptscriptstyle\rightharpoonup$}}
\over a}$ is the total acceleration perturbation difference between the satellites,
$\mathrm{\Delta} \mathord{\buildrel{\lower3pt\hbox{$\scriptscriptstyle\rightharpoonup$}}
\over g}$ and $\mathrm{\Delta}
{\mathord{\buildrel{\lower3pt\hbox{$\scriptscriptstyle\rightharpoonup$}} \over a} _{ng}}$ are
the gravitational acceleration perturbation difference and non-gravitational one,
respectively. The total acceleration perturbation difference $\mathrm{\Delta}
\mathord{\buildrel{\lower3pt\hbox{$\scriptscriptstyle\rightharpoonup$}} \over a}$ can be
determined from the range-rate measurements based on the perturbation theory. The
gravitational accelerations are the strongest forces acting on the satellites and mainly
determine the orbits, which are directly related to the distance between two satellites. In
this study we focus on the effects of measurement errors on the Earth＊s gravity field
recovery, therefore we ignore the accelerations acting on the low-flying satellite caused by
the third bodies, such as the Moon, the Sun and other celestial bodies. The non-gravitational
accelerations also have a significant impact on the satellite which are measured by an
on-board accelerometer, although they are smaller than the gravitational ones. The
acceleration perturbation difference with respect to local orbital reference frames shown in
Fig. 1, which correspond to the along-track, cross-track and radial directions of each
satellite \citep{Mackenzie and Moore 1997}, are provided as follows:
\begin{eqnarray} \label{eq2}
\left\{ {\begin{array}{*{20}{c}}
{\Delta {a_x} = \Delta {g_x} + \Delta {a_{x,ng}}}\\
{\Delta {a_y} = \Delta {g_y} + \Delta {a_{y,ng}}}\\
{\Delta {a_z} = \Delta {g_z} + \Delta {a_{z,ng}}}
\end{array}} \right.,
\end{eqnarray}
where $\mathrm{\Delta} {a_x}$, $\mathrm{\Delta} {a_y}$ and $\mathrm{\Delta} {a_z}$ are the
total acceleration perturbation differences in the along-track, cross-track and radial
directions, respectively; $\mathrm{\Delta} {g_x}$, $\mathrm{\Delta} {g_y}$ and
$\mathrm{\Delta} {g_z}$ are the gravitational ones; $\mathrm{\Delta} {a_{x,ng}}$,
$\mathrm{\Delta} {a_{y,ng}}$ and $\mathrm{\Delta} {a_{z,ng}}$ are non-gravitational ones. In
Fig. 1 $\eta$ and $\rho$ are the satellite separation and the intersatellite distance. Under
the assumption of a perfect polar circular orbit in this study, the local north-oriented
coordinate system is the same as the local orbital coordinate system. The range-rate
perturbations come from the along-track and radial perturbation difference, while cross-track
perturbation does not show up in this configuration with both satellites flying on the same
nominal orbit \citep{Sneeuw 2000}. As a result, we shall deal with the along-track and radial
components in this study. For the sake of clarity, Eq. \eqref{eq2} is rewritten as
\begin{eqnarray} \label{eq3}
\left\{ {\begin{array}{*{20}{c}}
{\Delta {g_x} = \Delta {a_x} - \Delta {a_{x,ng}}}\\
{\Delta {g_z} = \Delta {a_z} - \Delta {a_{z,ng}}}
\end{array}} \right.,
\end{eqnarray}
From Eq. \eqref{eq3} it can be seen that the accuracy of the retrieved gravitational
accelerations, which are the first order derivative of the gravitational potential, depends
on the ranging system and accelerometer noises. The next section presents the relationship
between measurement noises and the accuracies of gravity field coefficients based on the
above acceleration equations.

\section{Measurement noise and accuracies of gravity field coefficients}
\label{sec:3} The disturbance gravitational potential of the Earth is a harmonic function and
can be expanded into a series of spherical harmonics, depending on the spherical coordinates
$r$, $\theta$ and $\lambda$ \citep{Heiskanen and Moritz 1967}
\begin{eqnarray} \label{eq4}
T(r,\theta,\lambda) & = & \frac{GM}{R}\sum_{l=2}^\infty\left(\frac{R}{r}\right)^{l+1}
\sum_{m=0}^l
\bar{P}_{lm}\left(\cos{\theta}\right)\times
                                \nonumber\\
& & {}\left(\bar{C}_{lm}\cos{m\lambda}+
\bar{S}_{lm}\sin{m\lambda}\right),
\end{eqnarray}
where

\begin{tabular}{l}
 $r, \theta, \lambda$ are geocentric spherical coordinates (radius, co-\\latitude, longitude);\\
 $R$ is reference length (mean semi-major axis of the Ea-\\rth); \\
 $GM$ is gravitational constant times mass of the Earth; \\
 $l, m$ are degree, order of spherical harmonic; \\
 $\bar{P}_{lm}\left(\cos{\theta}\right)$ are the fully normalized Legendre functi-\\ons, and result in the relation $\displaystyle\frac{ 1}{ 4\pi }\int _{\sigma }\bar{P}_{lm}(\cos \theta )^2\times$\\$\left(
\begin{array}{c}
 \cos m\lambda  \\
 \sin m\lambda
\end{array}
\right)^2  \mathrm{d}\sigma =1$, where $\displaystyle\int _{\sigma }\mathrm{d}\sigma$ means integration \\on the unit sphere; \\
 $\bar{C}_{lm}, \bar{S}_{lm}$ are fully normalized potential coefficients.
\end{tabular}
The range-rate between two satellites are the main measurements of LL-SST and related to the
gravitational potential difference along the orbit, which can be obtained based on Eq.
\eqref{eq4}
\begin{eqnarray} \label{eq5}
\begin{array}{l}
{T_{AB}} = {T_A}({r_A},{\theta _A},{\lambda _A}) - {T_B}({r_B},{\theta _B},{\lambda _B})\end{array},
\end{eqnarray}
\noindent where $T_{AB}$ is the difference of the disturbance gravitational potential between
the satellites. The gravitational acceleration difference in the $x$ and $z$ directions are
the first order derivatives of the gravitational potential difference in the corresponding
directions, and can be obtained by using the polar coordinates
\begin{eqnarray} \label{eq6}
\begin{array}{l}
T_{AB}^{\left( x \right)} = \mathrm{\Delta} {g_x}\\
\;\;\;\;\;\;\; = \displaystyle - \frac{1}{r}\frac{{\partial {T_{AB}}}}{{\partial \theta }},\\
\end{array}
\end{eqnarray}
\begin{eqnarray} \label{eq7}
\begin{array}{l}
T_{AB}^{\left( z \right)} = \mathrm{\Delta} {g_z}\\
\;\;\;\;\;\; = \displaystyle\frac{{\partial {T_{AB}}}}{{\partial r}},\\
\end{array}
\end{eqnarray}
where
\begin{displaymath}
\bar P_{lm}^{(\theta )}(\cos \theta ) = \frac{{d{{\bar P}_{lm}}(\cos \theta )}}{{d\theta }},
\end{displaymath}
and $r=r_{A}=r_{B}$ since the two satellites are the same orbit. The average powers of the
error of $T_{AB}^{\left( x \right)}$ and $T_{AB}^{\left( z \right)}$ over a sphere of radius
$r$ are
\begin{eqnarray} \label{eq8}
\sigma _{T_{AB}^{\left( x \right)}}^2 = \frac{1}{{4\pi }}\int_\sigma  {{{\left( {\delta T_{AB}^{\left( x \right)}} \right)}^2}d\sigma },
\end{eqnarray}
\begin{eqnarray} \label{eq9}
\sigma _{T_{AB}^{\left( z \right)}}^2 = \frac{1}{{4\pi }}\int_\sigma  {{{\left( {\delta T_{AB}^{\left( z \right)}} \right)}^2}d\sigma },
\end{eqnarray}
where $\delta T_{AB}^{\left( x \right)}$ and $\delta T_{AB}^{\left( z \right)}$ are the
errors of $T_{AB}^{\left( x \right)}$ and $T_{AB}^{\left( z \right)}$, respectively. As can
be seen from Eq. \eqref{eq3} they are caused by the measurement noises of total accelerations
and non-gravitational ones. Owing to the fact that $T_{AB}^{\left( z \right)}$ is also
expanded into a series of spherical harmonics, we can obtain the power of the errors of
gravitational acceleration difference in the $z$ direction by applying the orthogonality
property of spherical harmonics and Parseval's theorem \citep{Colombo 1981}
\begin{eqnarray} \label{eq10}
\begin{array}{l}
\sigma _{T_{AB}^{\left( z \right)}}^2 = 2{\left( {\displaystyle\frac{{GM}}{{{R^2}}}} \right)^2}\sum\limits_{l = 2}^\infty  {{{\left( {l + 1} \right)}^2}{{\left( {\displaystyle\frac{R}{r}} \right)}^{2\left( {l + 2} \right)}} \times  } \\
\;\;\;\;\;\;\;\;\;\;\;\;\displaystyle\sum\limits_{m = 0}^l {\left( {\sigma _{{{\bar C}_{lm}}}^2 + \sigma _{{{\bar S}_{lm}}}^2} \right)},
\end{array}
\end{eqnarray}
where $\sigma _{{{\bar C}_{lm}}}^2$ and $\sigma _{{{\bar S}_{lm}}}^2$ are the error variances of corresponding spherical harmonics. It is indicated that the uncertainties of spherical coefficients depend on the errors of the gravitational acceleration, which are caused by the noise from ranging system and accelerometer. For a specific value of $l$, the summation over $l$ at the right-hand side of Eq. \eqref{eq10} is removed and $\sigma _{T_{AB}^{\left( z \right)}}^2$ at left-hand side is updated with error degree power of gravitational accelerations in the $z$ direction $\sigma _{T_{AB}^{\left( z \right)},l}^2$, which represents the error power introduced in the $l$-th degree. Thus, error degree amplitudes ${\sigma _l}$, namely the square root of the error power of a certain degree, is obtained as follows:
\begin{eqnarray} \label{eq11}
\begin{array}{l}
{\sigma _l} = \sqrt {\sum\limits_{m = 0}^l {\left( {\sigma _{{{\bar C}_{lm}}}^2 + \sigma _{{{\bar S}_{lm}}}^2} \right)} } \\
\;\;\; \;= \displaystyle\frac{{{\sigma _{T_{AB}^{\left( z \right)},l}}}}{{\sqrt 2 \displaystyle\frac{{GM}}{{{R^2}}}{{\left( {\frac{R}{r}} \right)}^{l + 2}}\left( {l + 1} \right)}},
\end{array}
\end{eqnarray}
where
\begin{displaymath}
{\sigma _{T_{AB}^{\left( z \right)}}} = \sqrt {\sum\limits_{l = 2}^\infty  {\sigma
_{T_{AB}^{\left( z \right)},l}^2} }.
\end{displaymath}
For the sake of legibility, the transformation coefficient from ${\sigma _{T_{AB}^{\left( z
\right)},l}}$ to $\sigma _l$ is defined as $B(l)$
\begin{eqnarray} \label{eq12}
B\left( l \right) = \frac{1}{{\sqrt 2 \displaystyle\frac{{GM}}{{{R^2}}}{{\left( {\displaystyle\frac{R}{r}} \right)}^{l + 2}}\left( {l + 1} \right)}},
\end{eqnarray}
which is a function of degree $l$. Then Eq. \eqref{eq11} is written as follows:
\begin{eqnarray} \label{eq13}
{\sigma _l} = B\left( l \right){\sigma _{T_{AB}^{\left( z \right)},l}}.
\end{eqnarray}
Likewise, there is a similar relationship between error degree power of gravitational accelerations in the $x$ direction $\sigma _{T_{AB}^{\left( x \right)},l}^2$ and error degree amplitudes ${\sigma _l}$
\begin{eqnarray} \label{eq14}
{\sigma _l} = A\left( l \right){\sigma _{T_{AB}^{\left( x \right)},l}},
\end{eqnarray}
where $A(l)$ is the transform coefficient from $\sigma _{T_{AB}^{\left( x \right)},l}$ to
${\sigma _l}$. The along-track gravitational acceleration difference is the directional
derivative in the $x$ direction which leads to a loss of orthogonality of spherical
harmonics, so we cannot directly compute $A(l)$ based on the orthogonality property of
spherical harmonics and Parseval's theorem. In this study $A(l)$ is derived by utilizing the
definition of spherical harmonics and the integration property of associated Legendre
functions (see the Appendix for more details)
\begin{eqnarray} \label{eq15}
\begin{array}{l}
A(l) = \displaystyle\frac{1}{{\left( {\displaystyle\frac{{GM}}{{rR}}} \right){{\left( {\displaystyle\frac{R}{r}} \right)}^{l + 1}}}}\left[ {l\left( {l + 1} \right) - } \right.\\
\;\;\;\;\;\;\;\;\;\;\;{\left. {l(l + 1){P_l}(\cos \Delta \theta ) + P_l^2(\cos \Delta \theta )} \right]^{ - \frac{1}{2}}}.
\end{array}
\end{eqnarray}

As mentioned above, the errors of gravitational acceleration difference stem from
intersatellite ranging errors and non-gravitational forces errors, which are due to the
ranging system intrinsic and accelerometer noise, respectively. In order to investigate how
the intersatellite ranging errors degrade the accuracy of the gravity field recovery, we need
to establish the relationship between the range-rate perturbations and perturbing forces. In
this study, we define the transfer function from the range-rate perturbation $\delta \dot
\rho $ to the perturbing accelerations $\delta a_{x}$ and $\delta a_{z}$ as ${H_{\delta \dot
\rho \to \delta {a_x}}}$ and ${H_{\delta \dot \rho  \to \delta {a_z}}}$, respectively
(details will be discussed in Sect. 4). Under the hypothesis that the range-rate
perturbations are stationary stochastic noise, one obtains the PSDs of the perturbing
accelerations in the $x$ and $z$ directions caused by the range-rate perturbation, denoted as
${S_{\delta {a_x}}}(f)$ and ${S_{\delta {a_z}}}(f)$ (unit: $\mathrm{m/s^{2}/\sqrt{Hz}}$),
respectively
\begin{eqnarray} \label{eq17}
{S_{\delta {a_x}}}(f) = {S_{\delta \dot \rho }}(f){H_{\delta \dot \rho  \to \delta {a_x}}}(f),
\end{eqnarray}
\begin{eqnarray} \label{eq18}
{S_{\delta {a_z}}}(f) = {S_{\delta \dot \rho }}(f){H_{\delta \dot \rho  \to \delta {a_z}}}(f),
\end{eqnarray}
where ${S_{\delta \dot \rho }}(f)$ (unit: $\mathrm{m/s/\sqrt{Hz}}$) is the PSD of the noise
of range-rate measurements. Based on the definition of PSD and the relationship between
temporal frequencies and spherical harmonics \citep{Cai 2013a}, the error degree powers of
the perturbing accelerations ${\sigma _{\delta {a_x},l}}$ and ${\sigma _{\delta {a_z},l}}$,
which describe the error average power of the ones introduced by the range-rate errors in the
$l$-th degree, can be obtained as follows:
\begin{eqnarray} \label{eq21}
\begin{array}{l}
{\sigma _{\delta {a_x},l}} = \sqrt {\displaystyle\sum\limits_j {{{\int_{{f_j} - \mathrm{\Delta} f/2}^{{f_j} + \mathrm{\Delta} f/2} {\left[ {{S_{\delta {a_x}}}(f)} \right]} ^2}}df} } \\
{\rm{      }}\;\;\;\;\;\;\;\;\;\;\; = \sqrt {\displaystyle\sum\limits_j {{{\int_{{f_j} - \mathrm{\Delta} f/2}^{{f_j} + \mathrm{\Delta} f/2} {\left[ {{S_{\delta \dot \rho }}(f){H_{\delta \dot \rho  \to \delta {a_x}}}(f)} \right]} ^2}}df} }  ,
\end{array}
\end{eqnarray}
\begin{eqnarray} \label{eq22}
\begin{array}{l}
{\sigma _{\delta {a_z},l}} = \sqrt {\displaystyle\sum\limits_j {{{\int_{{f_j} - \mathrm{\Delta} f/2}^{{f_j} + \mathrm{\Delta} f/2} {\left[ {{S_{\delta {a_z}}}(f)} \right]}^2 }}df} } \\
{\rm{      }} \;\;\;\;\;\;\;\;\;\;\;= \sqrt {\displaystyle\sum\limits_j {{{\int_{{f_j} - \mathrm{\Delta} f/2}^{{f_j} + \mathrm{\Delta} f/2} {\left[ {{S_{\delta \dot \rho }}(f){H_{\delta \dot \rho  \to \delta {a_z}}}(f)} \right]}^2 }}df} }  ,
\end{array}
\end{eqnarray}
where$f_{j}$ are the spectral lines belongs to the $l$-th degree, and $\mathrm{\Delta}f$ is
the spectral resolution, which is determined by the spectral interval between two neighboring
spectral lines \citep{Cai  2013b}; e.g. when the length of a time-series is $T_{r}$, the
spectral resolution is $\mathrm{\Delta}f=1/T_{r}$. Based on the 2-D Fourier method and
modulation theorem, the spectral lines in each $l$-th degree spherical harmonics are
summarized as follows \citep[see][]{Cai  2013a}
\begin{displaymath}
f_j=\left\{
\begin{array}{l l}
{f_{l,0}},{f_{l,1}},{f_{l,2}}, \cdots {f_{l,l}}
 &  \textrm{(A)}\\[1mm]
{{f_{0,l - 1}},{f_{2,l - 1}},{f_{4,l - 1}}, \cdots {f_{l - 2,l - 1}}}
 &  \textrm{(B)}\\[1mm]
{{f_{0,l}},{f_{2,l}},{f_{4,l}}, \cdots {f_{l - 2,l}}}
 &  \textrm{(C)}
\end{array}
\right.
\end{displaymath}
for $l$ even, and
\begin{displaymath}
f_j=\left\{
\begin{array}{l l}
{f_{l,0}},{f_{l,1}},{f_{l,2}}, \cdots {f_{l,l}}
 &  \textrm{(A)}\\[1mm]
{{f_{1,l - 1}},{f_{3,l - 1}},{f_{5,l - 1}}, \cdots {f_{l - 2,l - 1}}}
 &  \textrm{(B)}\\[1mm]
{{f_{1,l}},{f_{3,l}},{f_{5,l}}, \cdots {f_{l - 2,l}}}
 &  \textrm{(C)}
\end{array}
\right.
\end{displaymath}
for $l$ odd, with
\begin{displaymath} \label{eq15}
f_{q,p}=\displaystyle \left(q+\frac{2p}{K_\lambda}\right)\cdot \rm{cpr},
\end{displaymath}
where $K_\lambda$ relates to the number of orbits and cpr is an abbreviation for one
cycle-per-revolution. The value of $f_{q,p}$ is mostly determined by $q$ since $K_\lambda$ is
a large number in reality. It is obvious that $f_{j}$ contains the spectral lines are close to 0,
2, 4, ..., $l-2$, $l$ cpr for $l$ even and 1, 3, 5, ..., $l-2$, $l$ cpr for $l$ odd, as shown
in Fig 2.
\begin{figure}
\begin{center}
\centering
% Use the relevant command to insert your figure file.
% For example, with the graphicx package use
  \includegraphics[angle=0, width=0.45\textwidth]{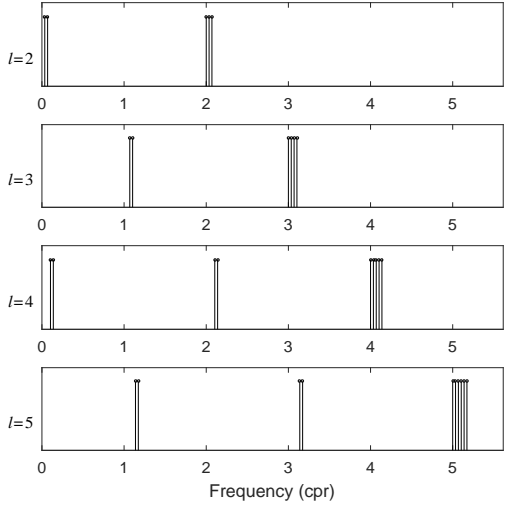}
% figure caption is below the figure
\caption{Spectral lines contained in $f_{j}$ for $l$=2, 3, 4, 5}
\label{fig:2}       % Give a unique label
\end{center}
\end{figure}
On the other hand, the noise level of accelerometer on board the satellite determines the
accuracies of the non-gravitational accelerations, and the error degree power of the
non-gravitational acceleration ${\sigma _{\delta {a_{ng}},l}}$, which describes the error
average power of the ones introduced by the satellites orbit errors in the $l$-th degree
\begin{eqnarray} \label{eq23}
{\sigma _{\delta {a_{ng}},l}} = \sqrt {\sum\limits_j {\int_{{f_j} - \mathrm{\Delta} f}^{{f_j} +
\mathrm{\Delta} f} {{{\left[ {{S_{acc}}(f)} \right]}^2}df} } },
\end{eqnarray}
where ${S_{acc}}(f)$ is the PSD of the accelerometer measurement noise.

On the basis of Eq. \eqref{eq3}, one can derive analytically the direct relationship between
the PSD of the LL-SST measurement errors and the coefficients of the Earth's gravity
potential by using the equations derived above. The direct relationship between the PSD of
range-rate errors and the coefficients of the Earth's gravity potential can be derived
analytically from Eqs. \eqref{eq13} and \eqref{eq14}:
\begin{eqnarray} \label{eq26}
\begin{array}{*{20}{l}}
{{\sigma _l} = A\left( l \right){\sigma _{T_{AB}^{\left( x \right)},l}}}\\
\begin{array}{l}
\;\;\;\; = A\left( l \right)\left\{ {\sum\limits_j {\left[ {\displaystyle\int_{{f_j} - \mathrm{\Delta} f/2}^{{f_j} + \Delta f/2} {{{\left( {{S_{\mathrm{\delta} \dot \rho }}(f){H_{\delta \dot \rho  \to {a_x}}}(f)} \right)}^2}} df + } \right.} } \right.\\
\;\;\;\;\;\;\;{\left. {\left. {\displaystyle\int_{{f_j} - \mathrm{\Delta} f}^{{f_j} + \mathrm{\Delta} f} {{{\left( {{S_{acc}}(f)} \right)}^2}df} } \right]} \right\}^{\frac{1}{2}}}
\end{array}
\end{array}
\end{eqnarray}
for the $x$ directions, and
\begin{eqnarray} \label{eq27}
\begin{array}{*{20}{l}}
{{\sigma _l} = B\left( l \right){\sigma _{T_{AB}^{\left( z \right)},l}}}\\
\begin{array}{l}
\;\;\;\; = B\left( l \right)\left\{ {\sum\limits_j {\left[ {\displaystyle\int_{{f_j} - \mathrm{\Delta} f/2}^{{f_j} + \Delta f/2} {{{\left( {{S_{\mathrm{\delta} \dot \rho }}(f){H_{\delta \dot \rho  \to {a_z}}}(f)} \right)}^2}} df + } \right.} } \right.\\
\;\;\;\;\;\;\;{\left. {\left. {\displaystyle\int_{{f_j} - \mathrm{\Delta} f}^{{f_j} + \mathrm{\Delta} f} {{{\left( {{S_{acc}}(f)} \right)}^2}df} } \right]} \right\}^{\frac{1}{2}}}
\end{array}
\end{array}
\end{eqnarray}
for the $z$ directions. In order to obtain the optimal solution, it is usual to recover the
gravity field from the combination of observations in the two directions, which is applied to
the following simulations.

\section{Relationship between perturbing forces and range-rate perturbations}
\label{sec:4} Using the orbit perturbation theory, this section describes the transfer
functions from range-rate perturbations to perturbing accelerations. Based on the assumption
of the polar circular orbit, the linearized Hill＊s equations are adopted here \citep{Colombo
1986,Schrama 1989}
\begin{eqnarray} \label{eq83}
\left\{ {\begin{array}{*{20}{c}}
{\ddot x + 2\omega \dot z = \delta {a_x}}\\
{\ddot y + {\omega ^2}y = \delta {a_y}}\\
{\ddot z - 2\omega \dot x - 3{\omega ^2}z = \delta {a_z}}
\end{array}} \right.,
\end{eqnarray}
where $\omega$ is the mean orbit rate $\omega  = \sqrt {GM/{r^3}} $, $\delta a_x$, $\delta
a_y$ and $\delta a_z$ are the perturbing accelerations in the along-track, cross-track and
radial directions, respectively. By applying the state space representation from control
system theory, the relationship between the perturbed state and the perturbing accelerations
can be expressed in the following state space form \citep{Kim 2000}:
\begin{eqnarray} \label{eq84}
\left\{ {\begin{array}{*{20}{c}}
{{\bf{\dot u}} = {\bf{Au}} + {\bf{Ba}}}\\
{{\bf{v}} = {\bf{Cu}}}
\end{array}} \right.,
\end{eqnarray}
where the perturbed state vector of two satellites is
\begin{displaymath}
{\bf{u}} = {\left[ {{x_1},{y_1},{z_1},{{\dot x}_1},{{\dot y}_1},{{\dot
z}_1},{x_2},{y_2},{z_2},{{\dot x}_2},{{\dot y}_2},{{\dot z}_2}} \right]^T},
\end{displaymath}
the perturbing acceleration vector
\begin{displaymath}
{\bf{a}} = {\left[ {{a_{{x_1}}},{a_{{y_1}}},{a_{{z_1}}},{a_{{x_2}}},{a_{{y_2}}},{a_{{z_2}}}}
\right]^T},
\end{displaymath}
and the perturbation vector of range-rate due to the perturbing forces
\begin{displaymath}
{\bf{v}} = {\left[ {\delta \rho ,\delta \dot \rho } \right]^T}.
\end{displaymath}
Accordingly, $\bf{A}$ and $\bf{B}$ are the coefficient matrices. Based on the Fig.
\ref{fig:1}, the intersatellite range-rate satisfy the following equations \citep{Visser
2005}:
\begin{eqnarray} \label{eq86}
\delta \dot \rho  = ({\dot z_2} + {\dot z_1})\sin \frac{\eta }{2} + ({\dot x_2} - {\dot x_1})\cos \frac{\eta }{2}.
\end{eqnarray}
Then $\bf{C}$ can be built in the following way:
\begin{displaymath}
{\bf{C}} = \left[ {\begin{array}{*{20}{c}} 0&0&0&{ - \cos \frac{\eta }{2}}&0&{\sin \frac{\eta
}{2}}&0&0&0&{\cos \frac{\eta }{2}}&0&{\sin \frac{\eta }{2}}
\end{array}} \right].
\end{displaymath}
The transfer function ${\bf{G}}(s)$, which maps the PSD of perturbing accelerations into that
of range-rate perturbations with zero initial conditions, can be computed analytically in the
complex frequency domain \citep{Ogata 2010}
\begin{eqnarray} \label{eq87}
\begin{array}{l}
{\bf{G}}(s) = {\bf{C}}{(s{\bf{I}} - {\bf{A}})^{ - 1}}{\bf{B}}\\
= \left[ {\begin{array}{*{20}{c}}
{{G_{{a_{{x_1}}} }}\!\!(s)}&{{G_{{a_{{y_1}}}}}\!\!(s)}&{{G_{{a_{{z_1}}}  }}\!\!(s)}&{{G_{{a_{{x_2}}}  }}\!\!(s)}&{{G_{{a_{{y_2}}}  }}\!\!(s)}&{{G_{{a_{{z_2}}} }}\!\!(s)}
\end{array}} \right]   .
\end{array}
\end{eqnarray}
In order to obtain the frequency transfer function ${\bf{G}}(f)$, the complex frequency $s$
is replaced by the frequency $f$ (where $s=2\pi f$). The transfer function ${\bf{H}}(f)$ from
the range-rate perturbations into perturbing accelerations is the reciprocal of ${\bf{G}}(f)$
based on their definitions \citep{Cai  2015}:
\begin{eqnarray} \label{eq88}
\begin{array}{l}
{\bf{H}}(f) \\
\!\!\!= \!\!\!\left[ {\begin{array}{*{20}{c}}
{{\!\!H_{{a_{{x_1}}}}}\!(f)}&{{\!H_{ {a_{{y_1}}}}}\!(f)}&{{\!H_{{a_{{z_1}}}}}\!(f)}&{{\!H_{{a_{{x_2}}}}}\!(f)}&{{\!H_{ {a_{{y_2}}}}}\!(f)}&{{\!H_{ {a_{{z_2}}}}}\!(f)}
\end{array}} \right]\\[2mm]
\!\!\!= \!\!\!\left[ {\begin{array}{*{20}{c}}
{\displaystyle\frac{1}{{{\!\!G_{{a_{{x_1}}}  }}\!(f)}}}&{\displaystyle\!\frac{1}{{{\!G_{{a_{{y_1}}} }}\!(f)}}}&{\displaystyle\!\frac{1}{{{\!G_{{a_{{z_1}}}  }}\!(f)}}}&{\displaystyle\!\frac{1}{{{\!G_{{a_{{x_2}}}  }}\!(f)}}}&{\displaystyle\!\frac{1}{{{\!G_{{a_{{y_2}}}  }}\!(f)}}}&{\displaystyle\!\frac{1}{{{\!G_{{a_{{z_2}}}  }}\!(f)}}}
\end{array}} \right].
\end{array}
\end{eqnarray}
Under the assumption of an orbit height of 450 km, the frequency response of transfers
function from the range-rate perturbations to perturbing forces can be obtained from the
above results, as shown in Fig. 3.
\begin{figure}
\begin{center}
\centering
% Use the relevant command to insert your figure file.
% For example, with the graphicx package use
  \includegraphics[angle=0, width=0.45\textwidth]{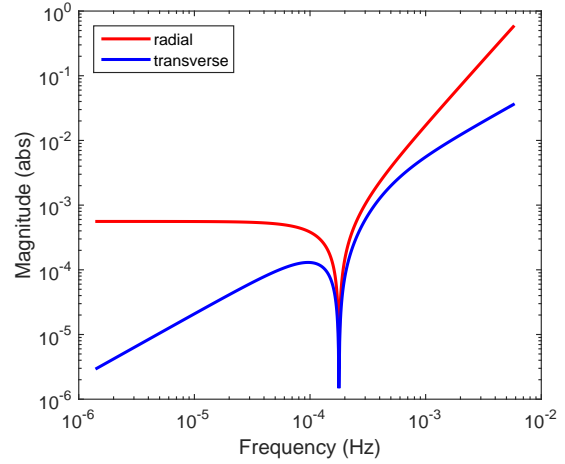}
% figure caption is below the figure
\caption{Transfer functions from range-rate perturbations to perturbing forces}
\label{fig:3}       % Give a unique label
\end{center}
\end{figure}
\section{Simulation and discussion}
\label{sec:5} The method enables us to gain a deeper insight into the error analysis of
LL-SST and is useful for mission design and error analysis. Considering the requirements of
GRACE-FO, we concentrate on the effects of instrument noises and orbit parameters on the
accuracy of the gravity field recovery.

\subsection{Sensor noise effects and matching}
\subsubsection{Noise realizations}
The realistic noises of onboard instruments, such as the intersatellite ranging instrument
and accelerometer, are generally colored. Based on a synthesis of the requirements from
\citet{Sheard 2012} and \citet{Elsaka 2014}, it is assumed that the PSD of a laser
interferometer ${S_{\dot \rho \_{\rm{LRI}}}}(f)$ is defined by means of the following
analytical functions:
\begin{eqnarray} \label{eq89}
{S_{\dot \rho \_{\rm{LRI}}}}(f) \!\!= \!\!2\pi f\sqrt {{{\left( {{S_0}} \right)}^2}\!\! +\!\!
\frac{{80}}{f}{{\left( {355 \!\!\times\!\! {{10}^{ - 12}} \cdot \frac{\rho }{{100\;{\rm{km}}}}}
\right)}^2}}
\end{eqnarray}
where ${\left( {{S_0}} \right)^2}$ and $80/f{\left( {355 \times {{10}^{ - 12}} \cdot \rho
/100\;{\rm{km}}} \right)^2}$ are the white noise component and frequency-dependent noise
component, respectively. $S_0$ is generally assigned a value of 50 $
\text{nm}/\sqrt{\textrm{Hz}}$ for LRI \citep{Elsaka 2014}. The factor $2\pi f$ relates to the
conversion of ranges to range-rates.

The main error sources of KBR onboard GRACE-FO are the oscillator and system noise. The PSD
of KBR noise model can be written as \citep{Kim 2000}
\begin{eqnarray} \label{eq90}
{S_{\dot \rho \_{\rm{KBR}}}}(f) = 2\pi f\sqrt {{{\left[ {{S_{{\rm{osc}}}}(f)} \right]}^2} + {{\left[
{{S_{{\rm{sys}}}}(f)} \right]}^2}},
\end{eqnarray}
where ${S_{{\rm{osc}}}}(f)$ and ${S_{{\rm{sys}}}}(f)$ are the PSD of the oscillator noise and
system noise, respectively. \citet{Kim 2000} describes the KBR measurement error due to the
oscillator and system noise following the GRACE case. The accelerometer noise model ACC 1 is
derived from the sensitive axes of a SuperSTAR-type sensor \citep{Touboul 1999}, which is the
accelerometer of GRACE (and expectedly also of GRACE-FO). The accelerometer noise contributes
are the detector, action, measure, parasitic and thermal noise \citep{Christophe 2010}.

In order to discuss the match between accelerometer and range-rate noise in spectral domain,
other three accelerometer noise models are introduced in this study, as shown in Table 1.
\begin{table}[htbp]
\caption{Four accelerometer noise models with different PSD}
\centering
\begin{tabular}{p{1cm}p{3.4cm}p{2.2cm}}
\toprule
Model & PSD (unit: ${\rm{m/}}{{\rm{s}}^2}/\sqrt{\textrm{Hz}})$  & Study (Ref.)\\
\cmidrule[0.5pt]{1-3}\\[-4mm]
ACC 1 & $1\!\!\times\!\! {10^{ -10}}\sqrt{1 \!\!+\!\! 0.005/f} $  & GRACE \& GRACE--FO  \citep{Kim 2000}\\ [2mm]
ACC 2 & $5\!\!\times\!\! {10^{ -11}}\sqrt{1 \!\!+\!\! 0.005/f} $  & e$^{2}$.motion \citep{Gruber 2014}\\  [2mm]
ACC 3 & $1.5\!\!\times\!\! {10^{ -12}}\sqrt{1 \!\!+\!\! 0.005/f} $  & NG2 \citep{Anselmi  2011}\\  [2mm]
ACC 4 & $1.5\!\!\times\!\! {10^{ -12}}$  & --- \\
\bottomrule
\end{tabular}
\end{table}
For comparison's sake, the analytic transfer function ${\bf{H}}(f)$ is applied in order to
convert range-rate perturbations into equivalent the accelerometer noise, and then the
accelerometer error is comparable to the range-rate errors. Figure 4 shows the PSD of the
accelerometer error due to KBR, LRI and different accelerometer noise models. The total noise
is dominated by the ACC noise at the low frequencies (with regard to GRACE-FO, $f<0.8$ mHz
for KBR, $f<6$ mHz for LRI), whereas by ranging system noise at the high frequencies.
Therefore, it is evident that the accuracy of the low degree gravity coefficients will be
mainly affected by the accelerometer noise, whereas the high degree ones be mainly affected
by the range system noise.
\begin{figure}
\begin{center}
\centering
% Use the relevant command to insert your figure file.
% For example, with the graphicx package use
  \includegraphics[angle=0, width=0.45\textwidth]{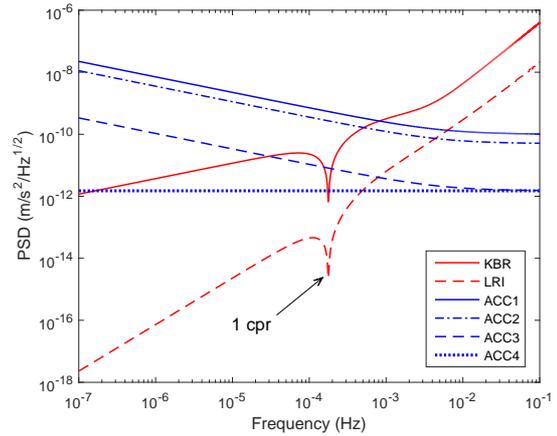}
% figure caption is below the figure
\caption{Comparison of the power spectra density of acceleration noises due to range-rate errors
 and different accelerometer noise models. The red and the dashed red lines denote the KBR noise and
 the LRI noise, respectively. The blue lines from top to bottom are accelerometer noises
 ACC 1, ACC 2, ACC 3 and ACC 4, respectively.
}
\label{fig:4}       % Give a unique label
\end{center}
\end{figure}

\subsubsection{Sensor noise effects}
The fundamental measurement quantity to be observed in a satellite to satellite tracking
mission are the distance variation between the two satellites and the non-gravitational
acceleration. This subsection, referring to GRACE-FO, discusses the differences in the
recovery caused from KBR and LRI measurement errors with the accelerometer noise ACC 1. For
this purpose, the following orbit parameters have been used for the simulation: orbit height
450 km; mission duration 12 months; separation distance 220 km. The PSD of instrument noise,
i.e. LRI, KBR and ACC 1, are above mentioned in the last subsection. According to the results
obtained in Sect. 3, the error degree amplitudes can be derived from the result of Sect. 3,
as shown in Fig. 5. Based on Kaula's rule, it can be concluded that the maximum recovery
degrees of the gravity field models are 197 and 160 for the LRI plus ACC1 and KBR plus ACC1,
corresponding to a half wavelength resolution of about 101 and 125 km, respectively. It is
obvious from Fig. 5 that the accuracy of the gravity field recovery recovery from LRI plus
ACC1 is improved about an order of magnitude better than that from KBR plus ACC1 in the
higher degrees ($l>10$). But in the lower degrees ($l<10$) the accuracy can not be improved
because accelerometer noise is dominant in this range. The corresponding cumulative geoid
height errors are shown in Fig. 6. From Fig. 6 and Table 2, it is seen that the two scenario
provide geoids with accuracies of $1.2 \times 10^{-2}$ and 4.3 cm at degree 150, and 7.4 and
10.2 cm at their maximum recovery degrees. The model derived from scenario LRI plus ACC 1 is
about 35 times better than that from KBR plus ACC 1 measurements except for the lower
degrees. It also can be seen from Eqs. \eqref{eq26} and \eqref{eq27} that an $N$ times better
range-rate accuracy yields about an $N$ times better gravity field model when other elements
remain the same.
\begin{figure}
\begin{center}
\centering
% Use the relevant command to insert your figure file.
% For example, with the graphicx package use
  \includegraphics[angle=0, width=0.45\textwidth]{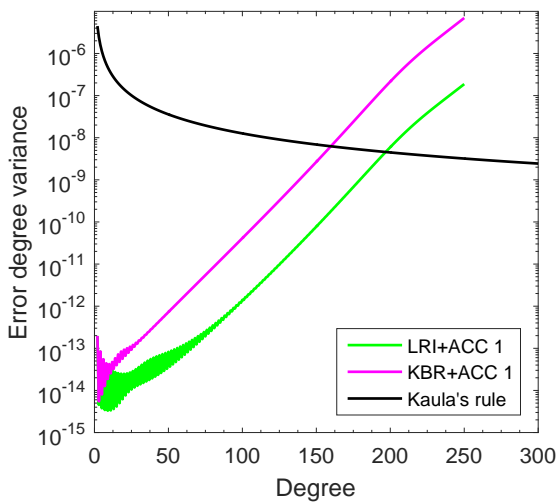}
% figure caption is below the figure
\caption{Error degree amplitudes under different scenarios}
\label{fig:5}       % Give a unique label
\end{center}
\end{figure}
\begin{figure}
\begin{center}
\centering
% Use the relevant command to insert your figure file.
% For example, with the graphicx package use
  \includegraphics[angle=0, width=0.45\textwidth]{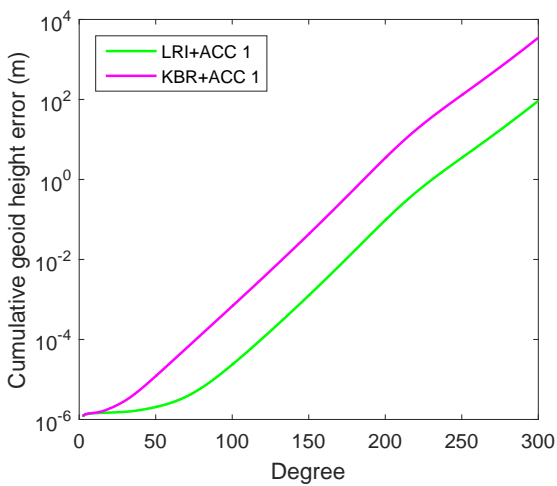}
% figure caption is below the figure
\caption{Cumulative geoid height errors under different scenarios}
\label{fig:6}       % Give a unique label
\end{center}
\end{figure}
\begin{table}[htb!]
\centering
\caption{Cumulative geoid height errors with different accuracies of range-rates}
\setlength{\tabcolsep}{2pt}
\begin{tabular}{p{2cm}p{1.0cm}p{1.6cm}p{1.6cm}p{1.4cm}p{0.1cm}}
\toprule
\multirow{2}{*}{ \tabincell{c}{Sensor noise\\ model}}&\multirow{2}{*}{ \tabincell{c}{Max. \\ degree}} & \multicolumn{4}{c}{Cumulative geoid height errors (cm)} \\
\cmidrule[0.5pt]{3-6}\\[-4mm]
&  & @Degree 100 &@Degree 150 &@Max. degree\\
\cmidrule[0.5pt]{1-6}   \\[-4mm]
LRI+ACC 1 & 197  & $2.4\times 10^{-3}$ & $1.2\times 10^{-1}$ & 7.4  \\
KBR+ACC 1 & 160  & $6.8\times 10^{-2}$ & 4.3                 & 10.2 \\
\bottomrule
\end{tabular}
\label{tab:wide_table1}
\end{table}

\subsubsection{Spectral matching of instrument accuracies}
The above results show that the accuracies of lower degree coefficients in GRACE-FO are
limited by the accelerometer noise, whether inter-satellite range-rate observations provided
by KBR or LRI. This subsection discusses the matching relation between range-rate
observations and accelerometer noises by taking advantage of the one-to-one correspondence
between spherical harmonics and frequencies indicated in the Sect. 3. For this purpose, the
following orbit parameters have been used for the simulation: orbit height 450 km; mission
duration 30 days; separation distance 220 km. Under these conditions, one can derive the
error degree amplitudes of the above-mentioned KBR, LRI and four accelerometer noise models,
as shown in Fig. 7. For the sake of clarity, we shall deal with the observations in the $x$
direction, although the observations in the $z$ direction has a similar property.

The comparison of the PSD and error degree amplitudes of sensor noises, i.e. Figs 4 and 7,
provides a valuable insight into the spectral matching of instrument accuracies. The PSD of
ACC 2 is two times better than that of ACC 1, but accelerometer noise is also dominant
compared to KBR error in the lower degrees ($l<8$). This is most obvious at degree 2 which
contains the frequencies close to zero frequency and 2 cpr. The error degree amplitudes of
ACC 1 are  about two orders of magnitude higher than that of KBR error at degree 2. It is
critical because the square of ACC 2 PSD is approximation of $1/f$ behavior below 5 mHz. For
this reason, the PSD of ACC 2 is about four orders of magnitude higher than that of KBR error
at the frequencies close to zero frequency but a mere two times higher at the frequencies
close to 2 cpr. Meanwhile, the other even degrees also have this effect due to the fact that
they contains the spectral lines close to zero cpr too. Certainly, the effect at lower
degrees caused by the $1/f$ behavior is more obvious than higher degrees since the number of
high frequencies increases with degrees. A solution to this problem depends on the sufficient
suppression of the $1/f$ noise of accelerometers.

In contrast, the error degree amplitudes at odd degrees are unaffected by the frequencies
close to zero frequency since they only contain the frequencies close to odd cpr. A
significant phenomenon is that error degree amplitudes of the models that contain a pink
noise, e.g. ACC 1, ACC 2 and ACC 3, are obviously a saw-tooth curve which fluctuates up and
down depending on parity of $l$, crests for $l$ even and troughs for $l$ odd. On the other
hand, thanks to the frequency trap of KBR at 1 cpr caused by ${\bf{H}}(f)$, the error degree
amplitudes of that at odd degrees are hardly affected by the noise around this frequency.
This is the reason why the error degree amplitudes of KBR are lower than that of ACC 2 at
degree 3 even if the PSD of them are equal at 3 cpr.  When applying ACC 3, which targets a
factor of 33 sensitivity improvement over ACC 2, the error degree amplitudes of KBR and
accelerometer noises match each other at degree 2. It means that the accuracies of spherical
harmonic coefficients at all degrees are determined by KBR noise in this situation, as shown
in Fig. 7. Considering the existing state-of-the-art accelerometer accuracy is at the level
of around ${10^{ -12}}$ ${\rm{m/}}{{\rm{s}}^2}/\sqrt{\textrm{Hz}}$ \citep{Drinkwater 2007},
it is necessary to make the accuracies of coefficients at lower degrees match each other by
removing or suppressing the $1/f$ noise of accelerometers if LR is applied instead of KBR.
Compared with the colored noise ACC 3, the white noise ACC 4 gets better match to LR at lower
degrees, as shown in Fig. 7. Based on the benefit of the elimination of $1/f$ noise, the
error degree amplitudes of ACC 4 are lower than that of ACC 3 at all degrees, although this
phenomenon decreases with degrees. For the same reason the saw-tooth behaviour of the error
degree amplitude curve disappears and then the accuracies of spherical harmonic coefficients
become more homogeneous between even and odd degrees. It is worth studying on the improvement
of the accuracy of accelerometers for the match between the range errors and accelerometer
noises in the future mission.
\begin{figure}
\begin{center}
\centering
% Use the relevant command to insert your figure file.
% For example, with the graphicx package use
  \includegraphics[angle=0, width=0.45\textwidth]{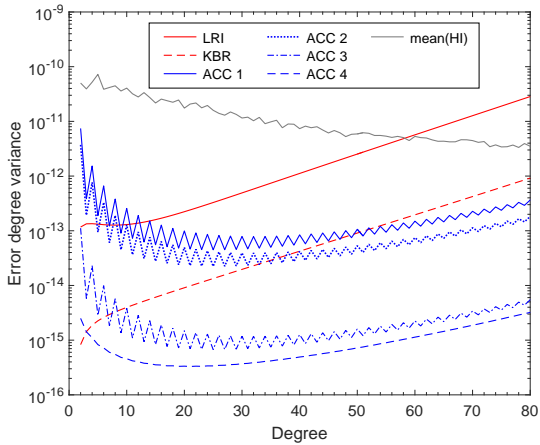}
% figure caption is below the figure
\caption{Comparison of error degree amplitudes of different sensor noises. The red and the
dashed red lines denote error degree amplitudes derived from the KBR and
 LRI noise, respectively. The blue lines from top to bottom are error degree amplitudes derived
 from accelerometer noises ACC 1, ACC 2, ACC 3 and ACC 4, respectively.}
\label{fig:7}       % Give a unique label
\end{center}
\end{figure}

\subsection{Time variable gravity signal effects}
The LL-SST missions can be effectively used for obtaining information on the temporal changes
of the Earth＊s gravity field on a global scale, which has been accompanied by temporal
aliasing due to undersampling of unmodeled mass variations \citep{Murbock 2014}. The method
proposed in this study can also be applied in the analysis of time variable gravity signal
effects. To investigate temporal aliasing caused by the time variable gravity signals, the
SST-ll observations are computed in terms of range-rate differences along the line-of-sight
of two satellites. The parameters of orbit are the same as stated already in the last
subsection except for the duration which is 30 days. There are two input time variable
gravity signals. The first signal is computed from the the residual AO signal (AO --
mean(AO)) from the ESA-AOHIS model \citep{Gruber 2011} and the second signal is computed from
the difference of the ocean tide models EOT08a \citep{Savcenko and Bosch 2008} and FES2004
\citep{Lyard 2006}.
\begin{figure}
\begin{center}
\centering
% Use the relevant command to insert your figure file.
% For example, with the graphicx package use
  \includegraphics[angle=0, width=0.45\textwidth]{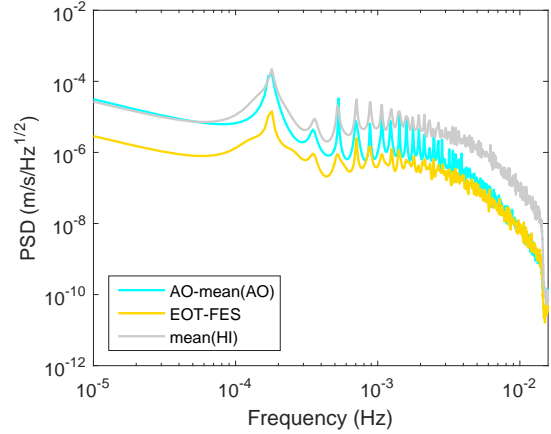}
% figure caption is below the figure
\caption{PSD of the difference of the non-tidal oceanic
and atmospheric signals and the corresponding mean signal and the difference of the
ocean tide models EOT08a and FES2004 together with the mean hydrology plus ice signal. They are
computed in terms of range-rate along the line-of-sight of two satellites}
\label{fig:8}       % Give a unique label
\end{center}
\end{figure}
Figure 8 shows the PSD of the two signals together with the mean hydrology plus ice signal
(mean(HI)) in terms of range-rate differences. It is found that time variable gravity signal
effects from the two sources acts on multiples of cpr. As previously mentioned in Sect. 3,
the spectral lines contained in $f_{j}$ are close to the multiples of cpr so that the error
degree amplitudes derived from the time variable gravity signals are mainly determined by the
peaks of their PSD. Figure 9 shows the error degree amplitudes of temporal aliasing from both
non-tidal and tidal sources including two types of sensor noise together with the mean
hydrology plus ice signal. From fig. 9, it can be seen that error degree amplitudes of the
residual AO signals intersect that of mean hydrology plus ice signal at degree 70,
corresponding to a half wavelength resolution of about 286 km. The error degree amplitudes of
the difference of the ocean tide models EOT08a and FES2004 are about one order of magnitude
lower than that of the residual AO signals. If KBR and ACC 1 are adopted as the sensor noises
then the maximum recovery degree of hydrology plus ice signal models is 58. In this case, the
temporal aliasing mainly determines the model accuracy at the degrees lower than 45, whereas
sensor noise mainly determines that at degrees from 45 to 58. If LRI and ACC 1 are adopted as
the sensor noises then the model accuracy are almost totally determined by the temporal
aliasing. As a result, temporal aliasing due to undersampling of unmodelled high frequency
mass variations will be one of the most serious problems for future gravity missions which
use high quality sensors \citep{Gruber 2014}.
\begin{figure}
\begin{center}
\centering
% Use the relevant command to insert your figure file.
% For example, with the graphicx package use
  \includegraphics[angle=0, width=0.45\textwidth]{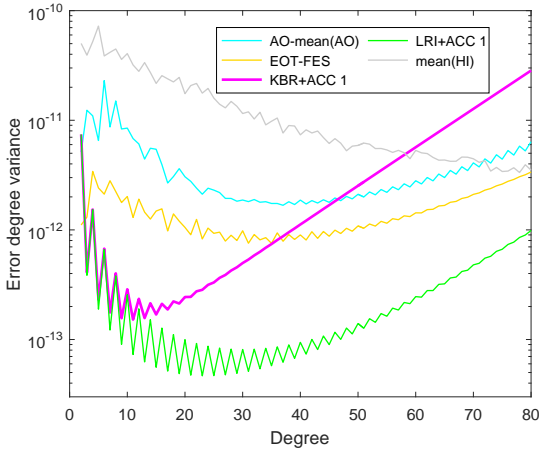}
% figure caption is below the figure
\caption{Error degree amplitudes of temporal aliasing from both non-tidal and tidal sources
        including two types of sensor noise together with the mean hydrology plus ice signal.}
\label{fig:9}       % Give a unique label
\end{center}
\end{figure}

\section{Conclusion}
Based on the spectral analysis and orbit perturbation theory, an analytical relationship
between the PSD of LL-SST measurements and the accuracies of gravity field coefficients is
presented mainly from the point of view of theory of signal and system, which indicates the
one-to-one correspondence between spherical harmonic error degree variances and frequencies
of the measurement noise. This relationship provides a physical insight into how the
measurement noises affect the accuracy of the gravity field recovery. The method is an
efficient and convenient tool for the design of future mission, especially for high accuracy
and resolution gravity field models. By taking GRACE-FO as the object of research, the
effects of sensor noises and time variable gravity signals are analyzed. If LRI measurements
are applied, a mission goal with a geoid accuracy of 7.4 cm at a spatial resolution of 101 km
is reachable, whereas if the KBR measurement error model is applied, a mission goal with a
geoid accuracy of 10.2 cm at a spatial resolution of 125 km is reachable. The spectral
matching of instrument accuracies is also investigated by taking the advantage of the
analytical relationship. It is necessary to improve the accuracy of accelerometers for the
match between the range errors and accelerometer noises in the future mission, especially for
removing or suppressing the 1/f noise. Temporal aliasing caused by the time variable gravity
signals is also discussed by this method. The one-to-one correspondence in the spectral
domain may provide a way for reducing the aliasing effects, but this still needs further
study based on the actual data.

This study is based on the hypothesis that the satellite orbit is a polar circular orbit,
while the realistic orbit with an inhomogeneous data distribution should cause a lower
accuracy and resolution model. It should be noted that the gravity signal can not exactly
recovered according to the Nyquist theorem if polar gaps occurs with a non-polar inclination.
In this case the results of error propagation computed by least-square methods are fitted
values, unless the gaps are filled with other data. Furthermore, the linear orbit
perturbation theory is adopted, which means that we have ignored the higher-order effect
terms. Notwithstanding its limits, the essential relationship is clearly indicated. Further
improvements in all these problems need to be further analyzed.

\begin{acknowledgements}
The authors are grateful to Prof. M. Zhong (IGG Wuhan) for his discussions. The valuable
suggestions and comments of Dr. L. Massotti (ESA), which improved the paper greatly, are
highly appreciated. This research is supported by the National Natural Science Foundations of
China under Grant No. 41404030 and 11235004.
\end{acknowledgements}

\section*{Appendix: Average power of the error of gravitational acceleration difference in the $x$
direction}
\setcounter{section}{1} \setcounter{subsection}{0} The expression for the average
power of gravitational acceleration difference errors is the key element for obtaining the
analytical relationship between the PSD of LL-SST measurements and the accuracies of gravity
field coefficients. Since the derivatives of the gravitational potential in the $z$ direction
keep the orthogonality property of spherical harmonics, it is easy to deduce the power of the
errors of gravitational acceleration difference by applying Parseval's theorem, as shown in
Eq. \eqref{eq10}. On the contrary, the derivatives of the gravitational potential in the $x$
direction relate to the derivatives with respect to co-latitude $牟$ and loose the
orthogonality property of spherical harmonics, so Parseval's theorem cannot be applied
directly in this situation. In this section, the average power of gravitational acceleration
difference errors in the $x$ direction is obtained based on the definition of spherical
harmonics and the integration property of associated Legendre functions.

According to Eq. \eqref{eq8}, the average power of gravitational acceleration difference
error in the $x$ direction can be expanded as
\begin{eqnarray} \label{eq28}
\begin{array}{l}
\sigma _{T_{AB}^{\left( x \right)}}^2\!\! = \!\!\displaystyle\frac{1}{{4\pi }}\int_\sigma  {{{\left( {T_{AB}^{\left( x \right)}} \right)}^2}d\sigma }  \\
\;\;\;\; \;\;\;\;\;\!\! = \!\!\displaystyle\frac{1}{{4\pi }}\!\!\iint_S {\left[ {{{\left( {\frac{{\partial {T_A}}}{{\partial x}}} \right)}^2} \!\!+ \!\!{{\left( {\frac{{\partial {T_B}}}
{{\partial x}}} \right)}^2} \!\!-\!\! 2\frac{{\partial {T_A}}}{{\partial x}}\frac{{\partial {T_B}}}{{\partial x}}} \right]}\times\\[4mm]
\;\;\;\; \;\;\;\;\;\;\;\sin \theta d\theta d\lambda   .
\end{array}
\end{eqnarray}
Substitution of Eq. \eqref{eq4} into above equation yields
\begin{eqnarray} \label{eq29}
\begin{array}{*{20}{l}}
{\sigma _{T_{AB}^{\left( x \right)}}^2\!\! =\!\! \displaystyle\frac{1}{{4\pi }}{{\left( {\frac{{GM}}{{rR}}} \right)}^2}\!\!\iint_S \!\!{\left\{ {\left[ {{{\sum\limits_{l = 2}^\infty  {\left( {\frac{R}{r}} \right)} }^{2\left( {l + 1} \right)}}\!\!\sum\limits_{m = 0}^l \!\!{\left( {{{\bar C}_{lm}}} \right. \times } } \right.} \right.} }\\
{\;\;\;\;\;\;\;\;\;{{\left. {\left. {\cos m{\lambda _A} + {{\bar S}_{lm}}\sin m{\lambda _A}} \right)\displaystyle\frac{{d{{\bar P}_{lm}}(\cos {\theta _A})}}{{d\theta }}} \right]}^2} \times }\\
\begin{array}{l}
\;\;\;\;\;\;\;\;\;\sin \theta d\theta d\lambda  + \left[ {{{\sum\limits_{l = 2}^\infty  {\left( {\displaystyle\frac{R}{r}} \right)} }^{2\left( {l + 1} \right)}}\sum\limits_{m = 0}^l {\left( {{{\bar C}_{lm}}} \right. \times } } \right.\\
\;\;\;\;\;\;\;\;\;{\left. {\left. {\cos m{\lambda _B} + {{\bar S}_{lm}}\sin m{\lambda _B}} \right)\displaystyle\frac{{d{{\bar P}_{lm}}(\cos {\theta _B})}}{{d\theta }}} \right]^2} \times \\
\;\;\;\;\;\;\;\;\;\sin \theta d\theta d\lambda  - 2\sum\limits_{l = 2}^\infty  {{{\left( {\displaystyle\frac{R}{r}} \right)}^{2\left( {l + 1} \right)}}\sum\limits_{m = 0}^l {\left[ {\left( {{{\bar C}_{lm}} \times } \right.} \right.} } \\
\;\;\;\;\;\;\;\;\;\left. {\left. {\cos m{\lambda _A} + {{\bar S}_{lm}}\sin m{\lambda _A}} \right)\displaystyle\frac{{d{{\bar P}_{lm}}(\cos {\theta _A})}}{{d\theta }}} \right] \times
\end{array}\\
\begin{array}{l}
\;\;\;\;\;\;\;\;\;\sum\limits_{l = 2}^\infty  {{{\left( {\displaystyle\frac{R}{r}} \right)}^{2\left( {l + 1} \right)}}\sum\limits_{m = 0}^l {\left[ {\left( {{{\bar C}_{lm}}\cos m{\lambda _B} + {{\bar S}_{lm}} \times } \right.} \right.} } \\
\;\;\;\;\;\;\;\;\;\left. {\left. {\sin m{\lambda _B}} \right)\displaystyle\frac{{d{{\bar P}_{lm}}(\cos {\theta _B})}}{{d\theta }}} \right] \times \;\sin \theta d\theta d\lambda \Bigg\}.
\end{array}\\
\end{array}
\end{eqnarray}
For a specific value of $l$, the summation over $l$ at the right-hand side of Eq.
\eqref{eq29} should be removed, and the error average power $\sigma _{T_{AB}^{\left( x
\right)}}^2$ at the left-hand side replaced with the error degree power $\sigma
_{T_{AB}^{\left( x \right)},l}^2$ , can be obtained as follows:
\begin{eqnarray} \label{eq30}
\begin{array}{*{20}{l}}
\begin{array}{l}
\sigma _{T_{AB}^{\left( x \right)},l}^2 = {\left( {\displaystyle\frac{{GM}}{{rR}}} \right)^2}{\left( {\displaystyle\frac{R}{r}} \right)^{2\left( {l + 1} \right)}}\left\{ {\displaystyle\frac{1}{{4\pi }}\iint\limits_S {\left[ {\sum\limits_{m = 0}^l {\left( {{{\bar C}_{lm}} \times } \right.} } \right.} } \right.\\
\;\;\;\;\;\;\;\;\;\;\;\;{\left. {\left. {\cos m{\lambda _A} + {{\bar S}_{lm}}\sin m{\lambda _A}} \right)\displaystyle\frac{{d{{\bar P}_{lm}}(\cos {\theta _A})}}{{d\theta }}} \right]^2} \times \\
\;\;\;\;\;\;\;\;\;\;\;\;\sin \theta d\theta d\lambda  + \displaystyle\frac{1}{{4\pi }}\iint\limits_S {\left[ {\sum\limits_{m = 0}^l {\left( {{{\bar C}_{lm}}\cos m{\lambda _B} + } \right.} } \right.} \\
\;\;\;\;\;\;\;\;\;\;\;\;{\left. {\left. {{{\bar S}_{lm}}\sin m{\lambda _B}} \right)\displaystyle\frac{{d{{\bar P}_{lm}}(\cos {\theta _B})}}{{d\theta }}} \right]^2}\sin \theta d\theta d\lambda  -
\end{array}\\
{\;\;\;\;\;\;\;\;\;\;\;\; \displaystyle\frac{2}{{4\pi }}\iint\limits_S {\sum\limits_{m = 0}^l {\bigg[\left( {{{\bar C}_{lm}}\cos m{\lambda _A} + {{\bar S}_{lm}}\sin m{\lambda _A}} \right) \times } } }\\
{\;\;\;\;\;\;\;\;\;\;\;\;\displaystyle\frac{{d{{\bar P}_{lm}}(\cos {\theta _A})}}{{d\theta }} \cdot \sum\limits_{m = 0}^l {\left( {{{\bar C}_{lm}}\cos m{\lambda _B} + {{\bar S}_{lm}} \times } \right.} }\\
{\;\;\;\;\;\;\;\;\;\;\;\;\left. {\left. {\sin m{\lambda _B}} \right)\displaystyle\frac{{d{{\bar P}_{lm}}(\cos {\theta _B})}}{{d\theta }}} \right]\sin \theta d\theta d\lambda \Bigg\} .}
\end{array}
\end{eqnarray}
Owing to the orthogonality of trigonometric functions, the summations over $l$ can be moved
outside of the square brackets, then one obtains
\begin{eqnarray} \label{eq31}
\begin{array}{*{20}{l}}
\begin{array}{l}
\sigma _{T_{AB}^{\left( x \right)},l}^2 = {\left( {\displaystyle\frac{{GM}}{{rR}}} \right)^2}{\left( {\displaystyle\frac{R}{r}} \right)^{2\left( {l + 1} \right)}}\left\{ {\displaystyle\frac{1}{{4\pi }}\sum\limits_{m = 0}^l {\displaystyle\iint\limits_S {\bigg[\left( {{{\bar C}_{lm}} \times } \right.} } } \right.\\
{\left. {\left. {\;\;\;\;\;\;\;\;\cos m{\lambda _A} + {{\bar S}_{lm}}\sin m{\lambda _A}} \right)\displaystyle\frac{{d{{\bar P}_{lm}}(\cos {\theta _A})}}{{d\theta }}} \right]^2} \times
\end{array}\\
\begin{array}{l}
\;\;\;\;\;\;\;\;\sin \theta d\theta d\lambda  + \displaystyle\frac{1}{{4\pi }}\displaystyle\sum\limits_{m = 0}^l {\iint\limits_S {\bigg[\left( {{{\bar C}_{lm}}\cos m{\lambda _B} + } \right.} } \\
\;\;\;\;\;\;\;\;{\left. {\left. {{{\bar S}_{lm}}\sin m{\lambda _B}} \right)\displaystyle\frac{{d{{\bar P}_{lm}}(\cos {\theta _B})}}{{d\theta }}} \right]^2}\sin \theta d\theta d\lambda  -
\end{array}\\
{\;\;\;\;\;\;\;\;\displaystyle\frac{2}{{4\pi }}\sum\limits_{m = 0}^l {\iint\limits_S {\bigg[\left( {{{\bar C}_{lm}}\cos m{\lambda _A} + {{\bar S}_{lm}}\sin m{\lambda _A}} \right) \times } } }\\
{\;\;\;\;\;\;\;\;\displaystyle\frac{{d{{\bar P}_{lm}}(\cos {\theta _A})}}{{d\theta }}\left( {{{\bar C}_{lm}}\cos m{\lambda _B} + {{\bar S}_{lm}} \times } \right.}\\
{\;\;\;\;\;\;\;\;\left. {\left. {\sin m{\lambda _B}} \right)\displaystyle\frac{{d{{\bar P}_{lm}}(\cos {\theta _B})}}{{d\theta }}} \right]\sin \theta d\theta d\lambda \} } .
\end{array}
\end{eqnarray}
The part between the brace in Eq. \eqref{eq31} consists of three integrals:

\noindent the first integral I1
\begin{eqnarray} \label{eq32}
\begin{array}{l}
{\rm{I}}1 = \displaystyle\frac{1}{{4\pi }}\sum\limits_{m = 0}^l {\iint_S {\left.\bigg[ {\left( {{{\bar C}_{lm}}\cos m{\lambda _A} + {{\bar S}_{lm}}\sin m{\lambda _A}} \right) \times } \right.} } \\
\;\;\;\;\;\;{\left. {\displaystyle\frac{{d{{\bar P}_{lm}}(\cos {\theta _A})}}{{d\theta }}} \right]^2}\sin \theta d\theta d\lambda,
\end{array}
\end{eqnarray}
the second integral I2
\begin{eqnarray} \label{eq33}
\begin{array}{l}
{\rm{I}}2 =\displaystyle \frac{1}{{4\pi }}\sum\limits_{m = 0}^l {\iint_S {\left.\bigg[ {\left( {{{\bar C}_{lm}}\cos m{\lambda _B} + {{\bar S}_{lm}}\sin m{\lambda _B}} \right) \times } \right.} } \\
\;\;\;\;\;\;{\left. {\displaystyle\frac{{d{{\bar P}_{lm}}(\cos {\theta _B})}}{{d\theta }}} \right]^2}\sin \theta d\theta d\lambda ,
\end{array}
\end{eqnarray}
and the third integral I3
\begin{eqnarray} \label{eq34}
\begin{array}{*{20}{l}}
\begin{array}{l}
{\rm{I}}3 = \displaystyle\frac{1}{{4\pi }}\sum\limits_{m = 0}^l {\iint\limits_S {\left.\bigg[ {\left( {{{\bar C}_{lm}}\cos m{\lambda _A} + {{\bar S}_{lm}}\sin m{\lambda _A}} \right) \times } \right.} } \\
\;\;\;\;\left( {{{\bar C}_{lm}}\cos m{\lambda _B} + {{\bar S}_{lm}}\sin m{\lambda _B}} \right)\displaystyle\frac{{d{{\bar P}_{lm}}(\cos {\theta _A})}}{{d\theta }} \times
\end{array}\\
{\left. {\;\;\;\;\displaystyle\frac{{d{{\bar P}_{lm}}(\cos {\theta _B})}}{{d\theta }}} \right]\sin \theta d\theta d\lambda }.
\end{array}
\end{eqnarray}
We only need to deal with two integrals since the first and second integrals are the same in
nature.

\subsection{Computation of the first and second integrals I1 \& I2}
In order to obtain of the first and second integral, we first compute it for a specific value
of $m$:
\begin{eqnarray} \label{eq35}
\begin{array}{l}
{\rm{I}}{1_m} = \displaystyle\frac{1}{{4\pi }}\iint_S {\left[ {\left( {{{\bar C}_{lm}}\cos m\lambda  + {{\bar S}_{lm}}\sin m\lambda } \right)} \right.} \\
\;\;\;\;\;\;\;{\left. {\displaystyle\frac{{d{{\bar P}_{lm}}(\cos \theta )}}{{d\theta }}} \right]^2}\sin \theta d\theta d\lambda ,
\end{array}
\end{eqnarray}
Considering the relationship between fully normalized Legendre polynomials and unnormalized
ones
\begin{eqnarray} \label{eq36}
{\bar P_{lm}}(\cos \theta ) = \sqrt {k(2l + 1)\frac{{(l - m)!}}{{(l + m)!}}}
{P_{lm}}(\cos\theta ),
\end{eqnarray}
where
\begin{displaymath}
k = \left\{ {\begin{array}{*{20}{c}}
{1\;\;\;{\rm{ for }}\;m = 0}\\
{2\;\;\;{\rm{ for }}\;m \ne 0}
\end{array}} \right.,
\end{displaymath}
${\rm{I}}{1_m}$ becomes
\begin{eqnarray} \label{eq37}
\begin{array}{l}
{\rm{I}}{1_m} = \displaystyle\frac{1}{{4\pi }}{\left( {\sqrt {k(2l + 1)\frac{{(l - m)!}}{{(l + m)!}}} } \right)^2} \times \\
\;\;\;\;\;\;\;\;\displaystyle\int\limits_0^\pi  {{{\left( {\displaystyle\frac{{d{P_{lm}}(\cos \theta )}}{{d\theta }}} \right)}^2}\sin \theta d\theta \displaystyle\int\limits_0^{2\pi } {\left( {\bar C_{lm}^2 \times } \right.} } \\
\;\;\;\;\;\;\;\;\left. {{{\cos }^2}m\lambda  + \bar S_{lm}^2{{\sin }^2}m\lambda } \right)d\lambda .
\end{array}
\end{eqnarray}
Substituting
\begin{eqnarray} \label{eq38}
\int_0^{2\pi } {{{\sin }^2}m\lambda d\lambda }  = \left\{ {\begin{array}{*{20}{c}}
{0\;(m = 0)}\\
{\pi \;(m \ne 0)}
\end{array}} \right.
\end{eqnarray}
and
\begin{eqnarray} \label{eq39}
\int_0^{2\pi } {{{\cos }^2}m\lambda d\lambda }  = \left\{ {\begin{array}{*{20}{c}}
{\pi \;\;\;(m \ne 0)}\\
{2\pi \;(m = 0)}
\end{array}} \right.
\end{eqnarray}
into Eq. \eqref{eq37} yields
\begin{eqnarray} \label{eq40}
\begin{array}{l}
{\rm{I}}{1_m} = \displaystyle\frac{1}{2}(2l \!+\! 1)\displaystyle\frac{{(l \!-\! m)!}}{{(l\! +\! m)!}}\sigma _{lm}^2\int_0^{2\pi } \!\!{{{\left( {\displaystyle\frac{{d{P_{lm}}(\cos \theta )}}{{d\theta }}} \right)}^2} \times } \\
\;\;\;\;\;\;\;\;\;\;\;\sin \theta d\theta,
\end{array}
\end{eqnarray}
where
\begin{displaymath}
\sigma _{lm}^2 = \left( {\sigma _{{{\bar C}_{lm}}}^2 + \sigma _{{{\bar S}_{lm}}}^2} \right).
\end{displaymath}
is the error degree-order variance. Since the first derivative of Legendre polynomials has a
recurrence property in the following form
\begin{eqnarray} \label{eq41}
\begin{array}{l}
\displaystyle\frac{{d{P_{lm}}(\cos \theta )}}{{d\theta }} = \frac{1}{2}\left[ {(l + m)(l - m + 1) \times } \right.\\
\;\;\;\;\;\;\;\;\;\;\;\;\;\;\;\;\;\;\;\;\;\;\;\left. {P_{l(m - 1)}^{}(\cos \theta ) - P_{l(m + 1)}^{}(\cos \theta )} \right].
\end{array}
\end{eqnarray}
Eq. \eqref{eq40} becomes
\begin{eqnarray} \label{eq42}
\begin{array}{*{20}{l}}
\begin{array}{l}
{\rm{I}}{1_m} = \left( {\displaystyle\frac{{\left( {2l + 1} \right)}}{2}\displaystyle\frac{{\left( {l - m} \right)!}}{{\left( {l + m} \right)!}}} \right)\sigma _{lm}^2\displaystyle\int\limits_0^\pi  {\displaystyle\frac{1}{4}\left\{ {\left[ {{{\left( {l + m} \right)}^2} \times } \right.} \right.} \\
\;\;\;\;\;\;\;{\left. {{{\left( {l - m + 1} \right)}^2}P_{l(m - 1)}^{}\left( {\cos \theta } \right)} \right]^2} - 2\left( {l + m} \right) \times
\end{array}\\
{\;\;\;\;\;\;\;\;\left( {l - m + 1} \right)P_{l(m - 1)}^{}\left( {\cos \theta } \right)P_{l(m + 1)}^{}\left( {\cos \theta } \right) + }\\
{\;\;\;\;\;\;\;\;{{\left[ {P_{l(m + 1)}^{}\left( {\cos \theta } \right)} \right]}^2}\bigg\} \sin \theta d\theta .}
\end{array}
\end{eqnarray}
Letting $x = \cos \theta $, then
\begin{eqnarray} \label{eq43}
\begin{array}{*{20}{l}}
\begin{array}{l}
{\rm{I}}{1_m} =  - \displaystyle\frac{1}{4}\left( {\displaystyle\frac{{\left( {2l + 1} \right)}}{2}\displaystyle\frac{{\left( {l - m} \right)!}}{{\left( {l + m} \right)!}}} \right)\sigma _{lm}^2\displaystyle\int\limits_{ - 1}^1 {\left\{ {\left[ {{{\left( {l + m} \right)}^2} \times } \right.} \right.} \\
\;\;\;\;\;\;\;{\left. {{{\left( {l - m + 1} \right)}^2}P_{l(m - 1)}^{}\left( x \right)} \right]^2} - 2\left( {l + m} \right) \times \\
\;\;\;\;\;\;\;\left( {l - m + 1} \right)P_{l(m - 1)}^{}\left( x \right)P_{l(m + 1)}^{}\left( x \right) +
\end{array}\\
{\left. {\;\;\;\;\;\;\;{{\left[ {P_{l(m + 1)}^{}\left( x \right)} \right]}^2}} \right\}{\rm{d}}x.}
\end{array}
\end{eqnarray}
Eq. \eqref{eq43} has three basic integrals:
\begin{displaymath}
\textrm{(A)}\;\;\int_{ - 1}^1 {{{\left[ {P_{l(m - 1)}^{}\left( x \right)} \right]}^2}dx},
\end{displaymath}
\begin{displaymath}
\textrm{(B)}\;\;\int_{ - 1}^1 {{{\left[ {P_{l(m + 1)}^{}\left( x \right)} \right]}^2}dx},
\end{displaymath}
\begin{displaymath}
\textrm{(C)}\;\;\int_{ - 1}^1 {P_{l(m + 1)}^{}\left( x \right)P_{l(m - 1)}^{}\left( x
\right)dx}.
\end{displaymath}
Based on the formula for computing the modulus of associated Legendre functions, one can
obtain the results of integrals A and B as follows:
\begin{eqnarray} \label{eq44}
\int_{ - 1}^1 {{{\left( {P_{l(m - 1)}^{}\left( x \right)} \right)}^2}dx}  = \frac{2}{{2l +
1}}\frac{{(l + m - 1)!}}{{\left( {l - m + 1} \right)!}},
\end{eqnarray}
\begin{eqnarray} \label{eq45}
\int_{ - 1}^1 {{{\left( {P_{l(m + 1)}^{}\left( x \right)} \right)}^2}dx}  = \frac{2}{{2l +
1}}\frac{{(l + m + 1)!}}{{\left( {l - m - 1} \right)!}}.
\end{eqnarray}
We resolve integral C with the definition of associated Legendre function. Substitution the
Rodrigues' formula
\begin{eqnarray} \label{eq46}
P_{lm}^{}(x) = \frac{{{{(1 - {x^2})}^{\frac{m}{2}}}}}{{{2^l}l!}}\frac{{{d^{l + m}}}}{{d{x^{l
+ m}}}}{\left( {{x^2} - 1} \right)^l}
\end{eqnarray}
into integral C yields
\begin{eqnarray} \label{eq47}
\begin{array}{l}
\begin{array}{*{20}{l}}
{\displaystyle\int\limits_{ - 1}^1 {{P_{l(m + 1)}}(x){P_{l(m - 1)}}(x)dx} }\\
{\; = \displaystyle\frac{1}{{{2^{2l}}l!l!}}\displaystyle\int\limits_{ - 1}^1 {{{\left( {1 - {x^2}} \right)}^{\frac{{m - 1}}{2} + \frac{{m + 1}}{2}}}\displaystyle\frac{{{d^{l + m - 1}}}}{{{d^{l + m - 1}}}}{{\left( {{x^2} - 1} \right)}^l} \times } }
\end{array}\\
\;\;\;\;\displaystyle\frac{{{d^{l + m + 1}}}}{{{d^{l + m + 1}}}}{\left( {{x^2} - 1} \right)^l}dx.
\end{array}
\end{eqnarray}
Letting $X = {x^2} - 1$, then
\begin{eqnarray} \label{eq48}
\begin{array}{l}
\displaystyle\int_{ - 1}^1 {{P_{l(m + 1)}}(x){P_{l(m - 1)}}(x)dx} \\
\;{\rm{ = }}\displaystyle\frac{{{{\left( { - 1} \right)}^m}}}{{{2^{2l}}l!l!}}\displaystyle\int_{ - 1}^1 {{X^m}\displaystyle\frac{{{d^{l + m - 1}}{X^l}}}{{d{x^{l + m - 1}}}}\displaystyle\frac{{{d^{l + m + 1}}{X^l}}}{{d{x^{l + m + 1}}}}} dx.
\end{array}
\end{eqnarray}
Before doing the integration, it is noted that all derivatives of the function ${X^m}$ up to
the $(m-1)$-th derivative have (${x^2} - 1$) as a factor, and are therefore zero at $x =  \pm
1$ . If we integrate Eq. \eqref{eq48} by parts we get
\begin{eqnarray} \label{eq49}
\begin{array}{l}
\displaystyle\int\limits_{ - 1}^1 {{P_{l(m + 1)}}(x){P_{l(m - 1)}}(x)dx} \\
\;\;\;\;{\rm{ = }}\displaystyle\frac{{{{\left( { - 1} \right)}^m}}}{{{2^{2l}}l!l!}}\displaystyle\int\limits_{ - 1}^1 {{X^m}\displaystyle\frac{{{d^{l + m + 1}}{X^l}}}{{d{x^{l + m + 1}}}}\displaystyle\frac{{{d^{l + m - 1}}{X^l}}}{{d{x^{l + m - 1}}}}} dx\\
\;\;\;\; = \displaystyle\frac{{{{\left( { - 1} \right)}^m}}}{{{2^{2l}}l!l!}}\left[ {\left. {{X^m}\displaystyle\frac{{{d^{l + m + 1}}{X^l}}}{{d{x^{l + m + 1}}}} \cdot \displaystyle\frac{{{d^{l + m - 2}}{X^l}}}{{d{x^{l + m - 2}}}}} \right|_{ - 1}^1} \right. \times \\
\;\;\;\;\;\;\;\left. { - \displaystyle\int\limits_{ - 1}^1 {\displaystyle\frac{{{d^{l + m - 2}}{X^l}}}{{d{x^{l + m - 2}}}}\displaystyle\frac{d}{{dx}}\left( {{X^m}\frac{{{d^{l + m + 1}}{X^l}}}{{d{x^{l + m + 1}}}}} \right)} dx} \right].
\end{array}
\end{eqnarray}
Owing to the condition just stated, the boundary term at the start is zero. We can continue
by integrating the remaining integral by parts with throwing away the boundary term until we
have done $(l+m-1)$ integrations. At this point one can obtain
\begin{eqnarray} \label{eq50}
\begin{array}{l}
\begin{array}{*{20}{l}}
\begin{array}{l}
\displaystyle\int\limits_{ - 1}^1 {{P_{lm}}(x){P_{l(m - 1)}}(x)dx} \\
\;\; = \displaystyle\frac{{{{\left( { - 1} \right)}^m}{{( - 1)}^{l + m - 1}}}}{{{2^{2l}}l!l!}}\displaystyle\int\limits_{ - 1}^1  {X^l}\displaystyle\frac{{{d^{l + m - 1}}}}{{d{x^{l + m - 1}}}}
\end{array}\\
{\;\;\;\;\;\left( {{X^m}\displaystyle\frac{{{d^{l + m + 1}}{X^l}}}{{d{x^{l + m + 1}}}}} \right)dx} .
\end{array}
\end{array}
\end{eqnarray}
Because that the largest power of $x$ in ${X^l}$ and ${X^m}$ is ${x^{2l}}$ and ${x^{2m}}$,
respectively, one can deduce that the item with the largest power of ${X^m}\frac{{{d^{l + m +
1}}{X^l}}}{{d{x^{l + m + 1}}}}$ is $\frac{{\left( {2l} \right)!}}{{\left( {l - m - 1}
\right)!}}{x^{l + m - 1}}$, and
\begin{eqnarray} \label{eq51}
\begin{array}{l}
\displaystyle\frac{{{d^{l + m - 1}}}}{{d{x^{l + m - 1}}}}\left( {{X^m}\displaystyle\frac{{{d^{l + m + 1}}{X^l}}}{{d{x^{l + m + 1}}}}} \right)\\ [4mm]
\;\; = \displaystyle\frac{{\left( {2l} \right)!}}{{\left( {l - m - 1} \right)!}}\left( {l + m - 1} \right)!.
\end{array}
\end{eqnarray}
Then Eq. \eqref{eq50} can be written as
\begin{eqnarray} \label{eq52}
\begin{array}{*{20}{l}}
\begin{array}{l}
\displaystyle\int\limits_{ - 1}^1  {P_{l(m + 1)}}(x){P_{l(m - 1)}}(x)dx\\
\;\;\; = \displaystyle\frac{{{{\left( { - 1} \right)}^m}{{( - 1)}^{l + m - 1}}}}{{{2^{2l}}l!l!}}\displaystyle\frac{{\left( {2l} \right)!}}{{\left( {l - m - 1} \right)!}}\left( {l + m - 1} \right)! \times
\end{array}\\
{\;\;\;\;\;\;\displaystyle\int\limits_{ - 1}^1 {{X^l}dx} }  .
\end{array}
\end{eqnarray}
The integral in Eq. \eqref{eq52} can be solved in the following form
\begin{eqnarray} \label{eq53}
\int_{ - 1}^1 {{X^l}dx}  = \int_{ - 1}^1 {{{\left( {{x^2}{\rm{ - }}1} \right)}^l}dx}  =
{\left( { - 1} \right)^l}\frac{{{{\left( {l!} \right)}^2}{2^{2l + 1}}}}{{\left( {2l + 1}
\right)!}},
\end{eqnarray}
so plugging this into Eq. \eqref{eq52} we find that
\begin{eqnarray} \label{eq54}
\begin{array}{*{20}{l}}
\begin{array}{l}
\displaystyle\int\limits_{ - 1}^1 {{P_{l(m + 1)}}(x){P_{l(m - 1)}}(x)dx(x)dx} \\
\;\; = \displaystyle\frac{{{{\left( { - 1} \right)}^m}{{( - 1)}^{l + m - 1}}}}{{{2^{2l}}l!l!}}\displaystyle\frac{{\left( {2l} \right)!}}{{\left( {l - m - 1} \right)!}} \times
\end{array}\\[1cm]
{\;\;\;\;\;\;\left( {l + m - 1} \right)!{{\left( { - 1} \right)}^l}\displaystyle\frac{{{{\left( {l!} \right)}^2}{2^{2l + 1}}}}{{\left( {2l + 1} \right)!}}}\\  [2mm]
{\;\; =  - \displaystyle\frac{{\left( {l + m - 1} \right)!}}{{\left( {l - m - 1} \right)!}}\displaystyle\frac{2}{{\left( {2l + 1} \right)}}} ﹝
\end{array}
\end{eqnarray}
Substitution of Eqs. \eqref{eq44}, \eqref{eq45} and \eqref{eq54} into Eq. \eqref{eq43} yields
\begin{eqnarray} \label{eq55}
\begin{array}{*{20}{l}}
{{\rm{I}}{1_m} = \displaystyle\frac{1}{4}\left( {\displaystyle\frac{{\left( {2l + 1} \right)}}{2}\displaystyle\frac{{\left( {l - m} \right)!}}{{\left( {l + m} \right)!}}} \right)\sigma _{lm}^2\left\{ {{{\left( {l + m} \right)}^2} \times } \right.}\\[4mm]
{\;\;\;\;\;\;{{\left( {l - m + 1} \right)}^2}\displaystyle\frac{2}{{2l + 1}}\displaystyle\frac{{(l + m - 1)!}}{{\left( {l - m + 1} \right)!}} - 2 \times }\\[4mm]
\begin{array}{l}
\;\;\;\;\;\;\left( {l + m} \right)\left( {l - m + 1} \right)\left[ { - \displaystyle\frac{{\left( {l + m - 1} \right)!}}{{\left( {l - m - 1} \right)!}}\displaystyle\frac{2}{{\left( {2l + 1} \right)}}} \right] + \;\\[4mm]
\;\;\;\;\;\;\left. {\displaystyle\frac{2}{{2l + 1}}\displaystyle\frac{{(l + m + 1)!}}{{\left( {l - m - 1} \right)!}}} \right\}
\end{array}\\[1cm]
{\;\;\;\;\;\; = \sigma _{lm}^2\left( {{l^2} + l - lm - \displaystyle\frac{m}{2}} \right)} .
\end{array}
\end{eqnarray}
Owing to the following relationship
\begin{eqnarray} \label{eq56}
{\rm{I}}1 = \sum\limits_m^{} {{\rm{I}}{1_m}}
\end{eqnarray}
and the fact that there are $(2l+1)$ linearly independent spherical harmonics in the $l$-th
degree, which relates to only one spherical harmonic ${P_l}\left( {\cos \theta } \right)$ for
$m=0$, and two spherical harmonics, i.e. ${P_{lm}}(\cos \theta )sin\left( {m\lambda }
\right)$ and ${P_{lm}}(\cos \theta )\cos \left( {m\lambda } \right)$, for $m=1, 2 ,..., l$,
one can compute the integral I1 as follows:
\begin{eqnarray} \label{eq57}
\begin{array}{l}
{\rm{I}}1 = {\left. {\sigma _{lm}^2\left( {{l^2} + l - lm - \displaystyle\frac{m}{2}} \right)} \right|_{m = 0}} + \\
\;\;\;\;\;\;\;\;2\sum\limits_{m = 1}^l {\sigma _{lm}^2\left( {{l^2} + l - lm - \displaystyle\frac{m}{2}} \right)}.
\end{array}
\end{eqnarray}
We assume that the error powers of these $(2l+1)$ spherical harmonics $\sigma _{lm}^2$ are
equal. This is reasonable because the temporal spectral lines of spherical harmonics of the
same degree are in close proximity and ones of different degrees are farther apart \citep{Cai
2013a}. Therefore, the effects of instrument noise on the error powers of spherical harmonics
in the same degree are nearly equal, especially for the white noise. Then, Eq. \eqref{eq57}
can be computed as
\begin{eqnarray} \label{eq58}
{\rm{I}}1 = \frac{1}{2}l\left( {1 + l} \right)\left( {2l + 1} \right)\sigma _{lm}^2.
\end{eqnarray}
Noticing that \citep{Rummel 1993}
\begin{eqnarray} \label{eq59}
\sigma _l^2 = \left( {2l + 1} \right)\sigma _{lm}^2,
\end{eqnarray}
then
\begin{eqnarray} \label{eq60}
{\rm{I}}1 = \frac{1}{2}l\left( {1 + l} \right)\sigma _l^2.
\end{eqnarray}
In the same way,
\begin{eqnarray} \label{eq61}
{\rm{I}}2 = \frac{1}{2}l\left( {1 + l} \right)\sigma _l^2.
\end{eqnarray}

\subsection{Computation of the third integral}
The integral I3 can be dealt with by applying the properties of the covariance function of
spherical function. First, we define a square integrable and analytical function
$f(\theta,\lambda)$ which is expanded in a series of spherical harmonics on the unit sphere
\begin{eqnarray} \label{eq62}
\begin{array}{l}
f(\theta ,\lambda ) = \sum\limits_{l = 0}^\infty  {\sum\limits_{m = 0}^l {{{\bar P}_{lm}}(\cos \theta )\left( {{{\bar C}_{lm}}\cos m\lambda  + } \right.} } \\[4mm]
\;\;\;\;\;\;\;\;\;\;\;\;\;\;\left. {{{\bar S}_{lm}}\sin m\lambda } \right)  .
\end{array}
\end{eqnarray}
The covariance function of $f(\theta,\lambda)$ at points A and B can be presented as follows:
\begin{eqnarray} \label{eq63}
\begin{array}{*{20}{l}}
\begin{array}{l}
{\rm{Cov}}\left( {f(A),f(B)} \right)\\
\;\; = \displaystyle\frac{1}{{4\pi }}\!\!\displaystyle\iint\limits_S {\sum\limits_{l = 0}^\infty  {\sum\limits_{m = 0}^l {\left( {{{\bar C}_{lm}}\cos m{\lambda _A} \!\!+\!\! {{\bar S}_{lm}}\sin m{\lambda _A}} \right) \times } } }
\end{array}\\
{\;\;\;\;\;\left( {{{\bar C}_{lm}}\cos m{\lambda _B} + {{\bar S}_{lm}}\sin m{\lambda _B}} \right){{\bar P}_{lm}}(\cos {\theta _A}) \times }\\[2mm]
{\;\;\;\;\;{{\bar P}_{lm}}(\cos {\theta _B})\sin \theta d\theta d\lambda }   .
\end{array}
\end{eqnarray}
Swapping integration and summation order leads to
\begin{eqnarray} \label{eq64}
\begin{array}{*{20}{l}}
\begin{array}{l}
{\rm{Cov}}\left( {f(A),f(B)} \right)\\
\;\; = \displaystyle\frac{1}{{4\pi }}\!\!\sum\limits_{l = 0}^\infty  {\sum\limits_{m = 0}^l {\displaystyle\iint\limits_S {\left( {{{\bar C}_{lm}}\cos m{\lambda _A} \!\!+ \!\!{{\bar S}_{lm}}\sin m{\lambda _A}} \right) \times } } }
\end{array}\\
{\;\;\;\;\;\;\left( {{{\bar C}_{lm}}\cos m{\lambda _B} + {{\bar S}_{lm}}\sin m{\lambda _B}} \right){{\bar P}_{lm}}(\cos {\theta _A}) \times }\\[2mm]
{\;\;\;\;\;\;{{\bar P}_{lm}}(\cos {\theta _B})\sin \theta d\theta d\lambda } .
\end{array}
\end{eqnarray}
The covariance function of $f(\theta,\lambda)$ can be also expanded in a series of Legendre
polynomials \citep{Colombo 1981}
\begin{eqnarray} \label{eq65}
{\rm{Cov}}\left( {f(A),f(B)} \right) = \sum\limits_{l = 0}^\infty  {\sigma _l^2{P_l}\left( {\cos \psi } \right)},
\end{eqnarray}
where $\psi$ is the spherical distance between the two points. On the other hand, the
derivative with respect to $\theta$ can be moved outside of the summation in Eq. \eqref{eq34}
\begin{eqnarray} \label{eq66}
\begin{array}{*{20}{l}}
{{\rm{I}}3 =\displaystyle \frac{{{\partial ^2}}}{{\partial {\theta ^2}}}\left\{ {\displaystyle\frac{1}{{4\pi }}\sum\limits_{m = 0}^l {\left[ {\displaystyle\iint\limits_S {\left( {{{\bar C}_{lm}}\cos m{\lambda _A} + {{\bar S}_{lm}} \times } \right.} } \right.} } \right.}\\
{\;\;\;\;\;\;\left. {\sin m{\lambda _A}} \right){{\left( {{{\bar C}_{lm}}\cos m{\lambda _B} + {{\bar S}_{lm}}\sin m{\lambda _B}} \right)}_{lm}} \times }\\
{\left. {\left. {\;\;\;\;\;\;{{\bar P}_{lm}}(\cos {\theta _A}){{\bar P}_{lm}}(\cos {\theta _B})\sin \theta d\theta d\lambda } \right.\Bigg]}\Bigg\} \right.}  .
\end{array}
\end{eqnarray}
It is concluded that the value within the brace of Eq. \eqref{eq66}
 is $\sigma _l^2{P_l}\left( {\cos \psi } \right)$ by comparing Eqs. \eqref{eq64} and
 \eqref{eq65}. The integral I3 can be represented as
\begin{eqnarray} \label{eq67}
{\rm{I}}3 = \frac{{{\partial ^2}\left( {\sigma _l^2{P_l}\left( {\cos \psi } \right)} \right)}}{{\partial {\theta ^2}}}.
\end{eqnarray}
The spherical distance $\psi$ can be computed as follows \citep{Moritz 1972}:
\begin{eqnarray} \label{eq68}
\cos \psi  = \cos {\theta _A}\cos {\theta _B} + \sin {\theta _A}\sin {\theta _B}\cos \left( {{\lambda _B} - {\lambda _A}} \right).
\end{eqnarray}
Since the satellite orbit is a polar circular orbit, i.e. $\lambda_{A}=\lambda_{B}$, one can
obtain
\begin{eqnarray} \label{eq69}
\cos \psi  = \cos \left( {{\theta _A} - {\theta _B}} \right) = \cos \eta,
\end{eqnarray}
which means $\psi=\eta$, and
\begin{eqnarray} \label{eq70}
\begin{array}{*{20}{l}}
{\displaystyle\frac{{\partial f\left( {\theta ,\lambda } \right)}}{{\partial \theta }} = \displaystyle\frac{{\partial f\left( {\theta ,\lambda } \right)}}{{\partial \psi }}\displaystyle\frac{{\partial \psi }}{{\partial \theta }}}\\
\begin{array}{l}
\;\;\;\;\;\;\;\;\;\;\;\;\; = \displaystyle\frac{1}{{\sin \psi }}\displaystyle\frac{{\partial f\left( {\theta ,\lambda } \right)}}{{\partial \psi }}\left[ {\sin {\theta _A}\cos {\theta _B} - } \right.\\
\;\;\;\;\;\;\;\;\;\;\;\;\;\;\;\;\left. {\cos {\theta _A}\sin {\theta _B}\cos \left( {{\lambda _A} - {\lambda _B}} \right)} \right]
\end{array}\\
{\;\;\;\;\;\;\;\;\;\;\;\;\; = \displaystyle\frac{{\partial f\left( {\theta ,\lambda } \right)}}{{\partial \psi }},}
\end{array}
\end{eqnarray}
which means the partial derivatives with respect to $\psi$ and the ones to $\theta$ are
equal. Then one can obtain
\begin{eqnarray} \label{eq71}
{\rm{I}}3 = \frac{{{d^2}\left( {\sigma _l^2{P_l}\left( {\cos \psi } \right)} \right)}}{{d{\psi ^2}}}.
\end{eqnarray}
By applying the relationships \vspace{-30pt}
\begin{spacing}{2}
\begin{eqnarray} \label{eq72}
\begin{array}{l}
\displaystyle\frac{{d{P_l}\left( {\cos \theta } \right)}}{{d\theta }} = P_{l1}^{}\left( {\cos \theta } \right),\\
\displaystyle\frac{{dP_{l1}^{}\left( {\cos \theta } \right)}}{{d\theta }} = \frac{1}{2}\left[ {(l + 1)l{P_l}(\cos \theta ) - P_{l2}^{}(\cos \theta )} \right],
\end{array}
\end{eqnarray}
\end{spacing}
\vspace{-20pt} \noindent we get finally
\begin{eqnarray} \label{eq73}
{\rm{I}}3 = \frac{1}{2}\sigma _l^2\left[ {(l + 1)l{P_l}(\cos \eta ) - {P_{l2}}(\cos \eta )} \right].
\end{eqnarray}

It is should be pointed out that the integral I3 can be equivalent to the integral I1 and
integral I2 when points A and B are coincident. In this situation, the satellite separation
$\eta=0$, then $\cos \eta=1$ and
\begin{eqnarray} \label{eq74}
\begin{array}{l}
{P_l}(\cos \eta ) = 1,\\
{P_{l2}}(\cos \eta ) = 0 .
\end{array}
\end{eqnarray}
So plugging Eq. \eqref{eq74} into Eq. \eqref{eq73} we find that \vspace{-30pt}
\begin{spacing}{2}
\begin{eqnarray} \label{eq75}
\begin{array}{l}
\displaystyle{\rm{I}}1 = {\rm{I}}2 = \frac{1}{2}\sigma _l^2\left[ {(l + 1)l{P_l}(\cos \eta ) - {P_{l2}}(\cos \eta )} \right]\\
\displaystyle\;\;\;\;\;\;\;\;\;\;\;\; = \frac{1}{2}l(l + 1)\sigma _l^2 .
\end{array}
\end{eqnarray}
\end{spacing}
\vspace{-20pt} \noindent The above results are the same as those in last subsection.

\subsection{Results of computation}
The error degree power of gravitational acceleration difference in the $x$ direction is
obtained by substituting Eqs. \eqref{eq60}, \eqref{eq61} and \eqref{eq73} into Eq.
\eqref{eq31}
\begin{eqnarray} \label{eq76}
\begin{array}{l}
\sigma _{T_{AB}^{\left( x \right)},l}^2 = \sigma _l^2{\left( {\displaystyle\frac{{GM}}{{rR}}} \right)^2}{\left( {\displaystyle\frac{R}{r}} \right)^{2\left( {l + 1} \right)}}\left[ {l\left( {l + 1} \right) - } \right.\\
\;\;\;\;\;\;\;\;\;\;\;\left. {l(l + 1){P_l}(\cos \Delta \theta ) + P_l^2(\cos \Delta \theta )} \right] .
\end{array}
\end{eqnarray}
For the sake of clarity, Eq. \eqref{eq76} is rewritten as
\begin{eqnarray} \label{eq77}
\begin{array}{l}
\displaystyle\sigma _l^2 = \frac{{\sigma _{T_{AB}^{\left( x \right)},l}^2}}{{{{\left( {\displaystyle\frac{{GM}}{{rR}}} \right)}^2}{{\left( {\displaystyle\frac{R}{r}} \right)}^{2\left( {l + 1} \right)}}}}\left[ {l\left( {l + 1} \right) - } \right.\\
\displaystyle\;\;\;\;\;\;\;\;\;\;\;{\left. {l(l + 1){P_l}(\cos \Delta \theta ) + P_l^2(\cos \Delta \theta )} \right]^{ - 1}}.
\end{array}
\end{eqnarray}
We get the transform coefficient $A(l)$ by comparing Eqs. \eqref{eq14} with \eqref{eq77}
\begin{eqnarray} \label{eq80}
\begin{array}{l}
A(l) = \displaystyle\frac{1}{{\left( {\displaystyle\frac{{GM}}{{rR}}} \right){{\left( {\displaystyle\frac{R}{r}} \right)}^{l + 1}}}}\left[ {l\left( {l + 1} \right) - } \right.\\
\;\;\;\;\;\;\;\;\;\;\;{\left. {l(l + 1){P_l}(\cos \Delta \theta ) + P_l^2(\cos \Delta \theta )} \right]^{ - \frac{1}{2}}}.
\end{array}
\end{eqnarray}

%\section*{Appendix 2}
%\subsection{name 1}

%\subsection{name 2}

% BibTeX users please use one of
%\bibliographystyle{spbasic}      % basic style, author-year citations
%\bibliographystyle{spmpsci}      % mathematics and physical sciences
%\bibliographystyle{spphys}       % APS-like style for physics
%\bibliography{}   % name your BibTeX data base

% Non-BibTeX users please use

\end{document}